\documentclass[twocolumn,preprintnumbers,superscriptaddress,endnote,nofootinbib,aps,prd]{revtex4-1}
\pdfoutput=1

\usepackage[hypertex]{hyperref}

\usepackage{graphicx}
\usepackage{dcolumn}   % needed for some tables
\usepackage{amssymb}   % for math
\usepackage{amsmath}

\usepackage{colordvi}
\usepackage{epsfig}
\usepackage{latexsym}

\newcommand{ \centeron }[2]{{\setbox0=\hbox{#1}\setbox1=\hbox{#2}\ifdim
                             \wd1>\wd0\kern.5\wd1\kern-.5\wd0\fi \copy0
                             \kern-.5\wd0\kern-.5\wd1\copy1\ifdim\wd0>\wd1
                             \kern.5\wd0\kern-.5\wd1\fi}}
\newcommand{ \ltap }{\>\centeron{\raise.35ex\hbox{$<$}}
                     {\lower.65ex\hbox{$\sim$}}\>}
\newcommand{ \gtap }{\>\centeron{\raise.35ex\hbox{$>$}}
                     {\lower.65ex\hbox{$\sim$}}\>}

\newcommand{ \slashchar }[1]{\setbox0=\hbox{$#1$}   % set a box for #1
   \dimen0=\wd0                                     % and get its size
   \setbox1=\hbox{/} \dimen1=\wd1                   % get size of /
   \ifdim\dimen0>\dimen1                            % #1 is bigger
      \rlap{\hbox to \dimen0{\hfil/\hfil}}          % so center / in box
      #1                                            % and print #1
   \else                                            % / is bigger
      \rlap{\hbox to \dimen1{\hfil$#1$\hfil}}       % so center #1
      /                                             % and print /
   \fi}                                             %

\newcommand{\ETmiss}{\slashchar{E}_T}

\newcommand{\gev}{\text{GeV}}

\newcommand{\lsim}{\lesssim}
\newcommand{\gsim}{\gtrsim}
\newcommand{\ra}{\rightarrow}

\newcommand{\Lic}{\Lambda_{ic}}
\newcommand{\Nic}{N_{ic}}
\newcommand{\alphaic}{\alpha_{ic}}

\newcommand{\BR}{\mathrm{BR}}

\newcommand{\LQCD}{\Lambda_{\rm QCD}}
\newcommand{\alphas}{\alpha_s}
\newcommand{\alphaW}{\alpha_W}
\newcommand{\alphaem}{\alpha_{\rm em}}

\newcommand{\bV}{U}  % big V
\newcommand{\bS}{S}  % big S

\newcommand{\nzero}{u_0}
\newcommand{\none}{u_1}
\newcommand{\nni}{u_i}

\usepackage{pstricks}

\begin{document}

% page numbers bottom-center
\pagestyle{plain}

\preprint{FERMILAB-PUB-11-271-T}

\title{Quirks at the Tevatron and Beyond}

\author{Roni Harnik}
\affiliation{Theoretical Physics Department, Fermilab, Batavia, IL 60510}

\author{Graham D. Kribs}
\affiliation{Theoretical Physics Department, Fermilab, Batavia, IL 60510}
\affiliation{Department of Physics, University of Oregon,
             Eugene, OR 97403}

\author{Adam Martin}
\affiliation{Theoretical Physics Department, Fermilab, Batavia, IL 60510}

%%%%%%%%%%%%%%%%%%%%%%%%%%%%%%%%%%%%%%%%%%%%%%%%%%%%%%%%%%%%%%%%%%%%%%%%%%%%

\begin{abstract}

We consider the physics and collider phenomenology of quirks
that transform nontrivially under QCD color, $SU(2)_W$ as well as
an $SU(N)_{ic}$ infracolor group.  Our main motivation is to 
show that the recent $Wjj$ excess observed by CDF naturally arises 
in quirky models.  The basic pattern is that several different 
quirky states can be produced, some of which $\beta$-decay during 
or after spin-down, leaving the lightest electrically neutral quirks 
to hadronize into a meson that subsequently decays into gluon jets.  
We analyze LEP II, Tevatron, UA2, and electroweak precision constraints, 
identifying the simplest viable models:  scalar quirks (``squirks'') 
transforming as color triplets, $SU(2)_W$ triplets and singlets, 
all with vanishing hypercharge.  
We calculate production cross sections, weak decay, spin-down,
meson decay rates, and estimate efficiencies.
The novel features of our quirky model includes:
quirkonium decay proceeds into a pair of gluon jets, 
without a $b$-jet component;
there is essentially no associated $Z jj$ or $\gamma jj$ signal; and 
there are potentially new (parameter-dependent) contributions to 
dijet production, multi-$W$ production plus jets, $W\gamma$, 
$\gamma\gamma$ resonance signals, and monojet signals.
There may be either underlying event from low energy QCD deposition 
resulting from quirky spin-down, and/or qualitatively modified 
event kinematics from infraglueball emission.

\end{abstract}
\maketitle

%%%%%%%%%%%%%%%%%%%%%%%%%%%%%%%%%%%%%%%%%%%%%%%%%%%%%%%%%%%%%%%%%%%%%%%%%%%
\section{Introduction}
\label{sec:intro}

CDF has recently reported a $4.1\sigma$ excess in the 
$Wjj$ event sample in $7.3$~fb$^{-1}$ of data for dijet 
invariant masses between $120-160$~GeV \cite{Aaltonen:2011mk,CDFWjjwebsite}.  
The excess comprises of hundreds of events in the 
$\ell jj + \ETmiss$ channel, corresponding to a sizeable 
cross section, $\sigma \sim$~few~pb 
to account for the smaller leptonic branching fraction and the 
efficiency of detecting the signal.  
Several explanations 
have already appeared \cite{Eichten:2011sh,varyinginterest},
as well as detailed discussion of the size and shape
of the Standard Model (SM) background \cite{Campbell:2011gp}.

\begin{figure}[th!]
%\vspace{- 0.5cm}
\centering
\includegraphics[width=0.48\textwidth]{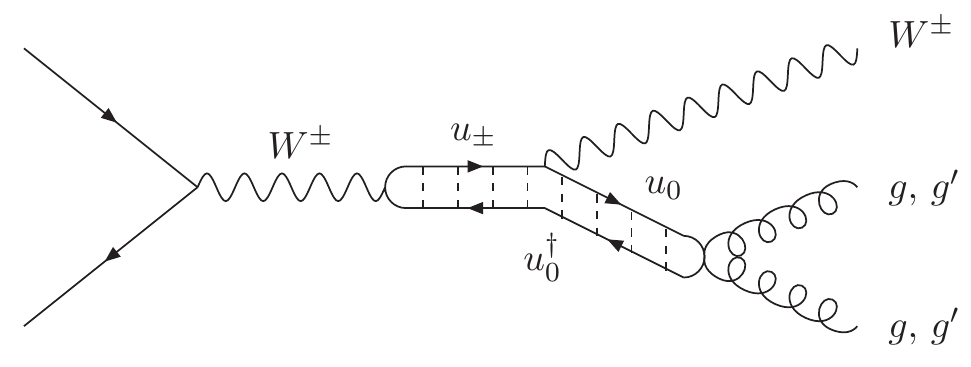}
\includegraphics[width=0.48\textwidth]{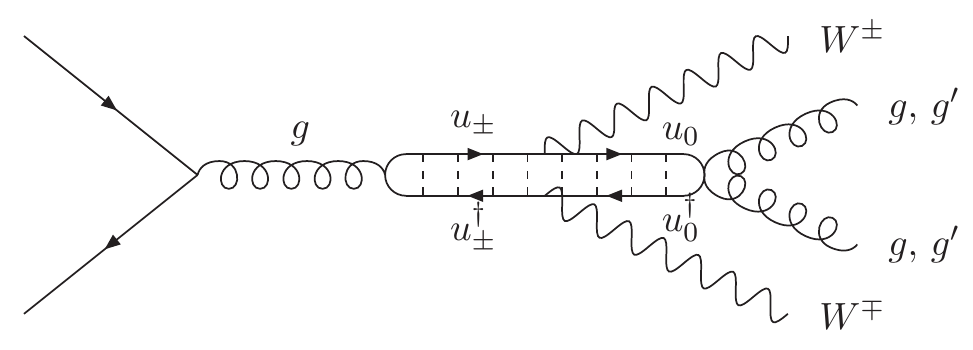}
\caption{Two classes of squirk pair production and $\beta$-decay
that can lead to a $Wjj$ final state with a dijet invariant mass peak
are shown in schematic diagrams.  The top, ``one-armed lobster'' 
diagram corresponds to weak production
$p\bar{p} \ra W^* \ra u_\pm \nzero^\dagger + \nzero u_\mp^\dagger$, 
following by charged squirk $\beta$-decay
$u_\pm \ra W^\pm \nzero$, 
and finally annihilation of the neutral quirks into a pair of gluons 
or infragluons ($g'$). 
The bottom, ``two-armed lobster'' diagram corresponds to 
strong production of a heavy quirk-anti-quirk pair, 
following by $\beta$-decay of both heavy squirks, and finally annihilation
of the neutral quirks into gluons. 
The latter production can pass the CDF analysis with a modest efficiency
when the decay products of one of the $W$'s are not detected.
Quirky spin-down radiation (into color gluons, infracolor glueballs, etc.) 
is not shown for either diagram.}
\label{fig:lobsters}
\end{figure}

In this paper we present an explanation of this excess invoking 
pair production of ``quirks'' \cite{Kang:2008ea}, defined below, 
which subsequently undergo radiative energy loss 
and weak decay, finally annihilating into dijets. 
There are two main pathways that begin with quirk production and end
with a $Wjj$ signal consistent with the CDF analysis after their cuts:
The first path consists of electroweak production of a heavy-light 
quirk pair, where the heavy quirk $\beta$-decays into a light quirk, 
emitting a $W^\pm$, and then the remaining quirk--anti-quirk system 
settles into a quirkonium ground state.  The quirkonium decays into 
gluon jets that reconstruct an invariant mass peak.
The second path consists of strong production of a heavy 
quirk-anti-quirk pair, where both quirks $\beta$-decay to their 
lighter electroweak partners.  Even though two $W$ bosons 
are emitted in this process, one $W$ decay can be missed 
by the CDF analysis, particularly when the decay products are
light quark jets that have energies below the CDF jet energy cuts. 
Schematic diagrams of the production 
and decay of the heavy-light quirk system and the heavy-heavy quirk 
system is shown in Fig.~\ref{fig:lobsters}.

%%%%%%%%%%%%%%%%%%%%%%%%%%%%%%%%%%%%
%%%%%%%%%%%%%%%%%%%%%%%%%%%%%%%%%%%%

In this paper we consider ``squirks'' -- the scalar analogues of 
``quirks'' \cite{old-quirks,Kang:2008ea}, 
which are new fields transforming under part of the SM gauge group 
along with a new strongly-coupled ``infracolor'' group $SU(N)_{ic}$.  
The strong coupling scale of the infracolor group, $\Lic$, 
is assumed to be much smaller than the masses of all quirky (or squirky)
fields transforming under infracolor.  Consequently, infracolor strings 
do not break, and quirk (or squirk) pairs remain in a bound state even 
when they are produced with high kinetic energies.
This leads to several interesting collider physics, model building 
and dark matter 
applications \cite{Kang:2008ea,Burdman:2006tz,Burdman:2008ek,Cheung:2008ke,Harnik:2008ax,Cai:2008au,Kilic:2009mi,Chang:2009sv,Nussinov:2009hc,Kribs:2009fy,Kilic:2010et,Carloni:2010tw,Martin:2010kk,FK2011,Luty-et-al}.
(Other work on hidden valley models can be found in 
\cite{Strassler:2006im,Han:2007ae,Strassler:2008fv}.)
Certain kinds of quirks have already been searched for 
at the Tevatron by the D0 collaboration \cite{Abazov:2010yb}.
Typically, squirks and quirks transform as a vector-like representation of
the SM gauge group as well as a vector-like 
representation of infracolor, and thus can acquire any mass without
any additional Higgs sector.\footnote{Noteable exceptions,
where quirks acquire masses through electroweak symmetry breaking,
can be found in Refs.~\cite{Kribs:2009fy,FK2011}.}  

Here is a summary of our basic strategy to explain the $Wjj$ signal,
while avoiding the many constraints from LEP II, Tevatron, and UA2:
\begin{itemize}
\item[1.] We consider squirks (scalars), rather than quirks (fermions), mainly
to employ a renormalizable operator to lead to mass splittings,
while also minimizing the electroweak precision contribution
resulting from this splitting.
\item[2.] The $Wjj$ signal arises from a combination of the two distinct
squirk pair production processes shown in Fig.~\ref{fig:lobsters}.
The weak decay, $u_\pm \ra W^\pm \nzero$, could be 2-body or 3-body, 
depending on the mass splitting between $u_\pm$ and $\nzero$.  
\item[3.] All of the squirks are color triplets under QCD\@.  
Hence, the leading visible annihilation channel of a squirk--anti-squirk
bound state system is into $gg$.  This provides our source of the 
jet-jet resonance, and consequently, we predict the flavor of the 
jets in the excess observed by CDF to be pure glue with no heavy flavor. 
\item[4.] The electrically-neutral squirks are mixtures among 
the $T_3 = 0$ component of an electroweak triplet as well as an
electroweak singlet.
Hence, they do not couple to the $\gamma$ or the $Z$.  
This implies the tree-level production cross section of the 
lightest squirks at LEP II vanishes.
\item[5.] The electrically-charged squirks have masses larger than
$100$~GeV, and so are beyond the LEP II sensitivity.
\item[6.] Light squirk pair production $\nzero \nzero^\dagger$ 
results in a dijet invariant mass around $150-160$~GeV\@.
The production cross section is far smaller than Tevatron's
dijet sensitivity \cite{Aaltonen:2008dn}, making it safe from 
Tevatron searches.  The production cross section of this resonance 
at the CERN SppS ($\sqrt{s} = 630$~GeV) collider is less than $1$~pb 
per infracolor.  The UA2 bound is roughly $90$~pb \cite{Alitti:1993pn}, 
and thus does not restrict $\Nic$.
\end{itemize}

At this point we should emphasize that some aspects of squirk production 
and decay can be calculated or simulated with standard 
collider tools, but some cannot and one must resort to estimates 
or parameterization of the various possibilities.  
What \emph{can} be calculated unambiguously is 
i) squirk pair production (at leading order, far from 
the cross section threshold),
ii) the weak decay rate of squirks (from which we use to 
infer the weak decay of the squirky mesons), and 
iii) squirky meson annihilation rates into
Standard Model fields.
What \emph{cannot} be calculated reliably is 
the dynamics of the 
``spin-down'' process, as the high energy squirks shed their 
momentum to settle into a (near) ground state squirky meson, 
and the spectrum and content of the resulting radiation.
An attempt at modeling this radiation is underway~\cite{Luty-et-al}.
As a consequence, we do not attempt a full simulation of 
squirky production. Instead, we calculate quantities 
that are under control, with estimates of signal efficiencies, 
and point out where simulations could improve our understanding.
To deal with the remaining uncertainties due to 
non-perturbative infracolor dynamics we discuss some of the 
possible scenarios for this spin down and their effects on phenomenology 
in Sec.~\ref{sec:spindowndecay}.

%%%%%%%%%%%%%%%%%%%%%%%%%%%%%%%%%%%%
\section{Model}
\label{sec:model}

We extend the Standard Model to include one set of squirks,
$\bV$ and $\bS$ with quantum numbers given in Table~\ref{tab:quantumnum}. 
%%%%%%%%%%%%%%%%%%%%%%%%%%%%%%%%%%%%
\begin{table}[t]
\begin{tabular}{c|cccc}
           & $SU(N)_{ic}$ & $SU(3)_c$    & $SU(2)_L$    & $U(1)_Y$ \\ \hline
$\bV$      & $\mathbf{N}$ & $\mathbf{3}$ & $\mathbf{3}$ &    0     \\
$\bS$      & $\mathbf{N}$ & $\mathbf{3}$ & $\mathbf{1}$ &    0     \\
\end{tabular}
\caption{Scalar quirk quantum numbers.}
\label{tab:quantumnum}
\end{table}
%%%%%%%%%%%%%%%%%%%%%%%%%%%%%%%%%%%%
The gauge-invariant operators involving these fields include 
the mass terms 
\begin{eqnarray}
\frac{1}{2} M_{\bV}^2 \bV^\dagger \bV + \frac{1}{2} M_{\bS}^2 \bS^\dagger \bS
\label{eq:vectormass}
\end{eqnarray}
and quartic interactions including
\begin{eqnarray}
& & \lambda_4 (\bS^\dagger \bS)(\bV^\dagger \bV)
+ \lambda_{\bV} (H^\dagger H)(\bV^\dagger \bV) 
+ \lambda_{\bS} (H^\dagger H)(\bS^\dagger \bS) \nonumber \\
& &{} + \lambda_{\bV 4} (U^\dagger U)^2 + \lambda_{\bS 4} (S^\dagger S)^2  
\label{eq:quartic}
\end{eqnarray}
as well as 
\begin{equation}
\kappa (H^\dagger \tau^a H)(\bS^\dagger \bV_a) + \mbox{h.c.} \; .
\label{eq:dim4}
\end{equation}
The operators proportional to $\lambda_{\bV,\bS}$ lead to shifts
in the masses of $\bV$ and $\bS$, as well as Higgs boson trilinear
and quartic interactions with $\bV^\dagger \bV$, $\bS^\dagger \bS$.
The operators proportional to $\lambda_{\bV 4, \bS 4}$ will turn out
to affect certain annihilation channels. 
Since our analysis does not depend on the existence of these
interactions, we do not consider them further.

The triplet field $\bV$ can be written in terms of its isospin components
\begin{eqnarray}
\bV &=& \left( \begin{array}{c} \bV_1 \\ \bV_2 \\ \bV_3 \end{array} \right) 
\end{eqnarray}
where we can rewrite the fields in the mass eigenstate basis 
with definite electric charges,
\begin{eqnarray}
u_\pm &\equiv& \frac{1}{\sqrt{2}} \left( \bV_1 \mp i \bV_2 \right) 
\end{eqnarray}
This is completely analogous to the rewriting of the usual triplet 
of $SU(2)_L$ of gauge bosons $W_{1,2,3}$ into $W^{\pm,0}$.

After electroweak symmetry breaking, the Higgs doublet can be
written as $H = (v + h) \exp[i \Pi]/\sqrt{2}$ in terms of the
vacuum expectation value $v \equiv 246$~GeV, the physical Higgs boson
$h$, and the nonlinear representation of the Goldstone bosons, 
contained within $\Pi$.  Expanding the operator Eq.~(\ref{eq:dim4}) 
to second order in $\Pi$, one can verify that only $(v+h)$ enters
the dynamics, and thus we can ignore the Goldstone interactions.

The central implication of Eq.~(\ref{eq:dim4}) is that it 
causes the singlet $\bS^\dagger$ to mix with the neutral component 
of the triplet $\bV_3$, enlarging the mass splitting between them.  
The two mass eigenstates, 
which we call $\nzero$ and $\none$ for the light and heavy states, 
have masses of 
\begin{eqnarray}
m_0^2 &=& \frac{1}{2} \left( m_{\bV}^2 +m_{\bS}^2 
              - \sqrt{(m_{\bV}^2-m_{\bS}^2)^2 + 4 \delta^4} \right) \\
m_1^2 &=& \frac{1}{2} \left( m_{\bV}^2 +m_{\bS}^2 
              + \sqrt{(m_{\bV}^2-m_{\bS}^2)^2 + 4 \delta^4} \right)
\end{eqnarray}
where $\delta \equiv \sqrt{\kappa v^2/2}$.
These neutral squirks are mixtures of the singlet and triplet
\begin{eqnarray}
\left( \begin{array}{c} \nzero \\ \none \end{array} \right) &=&
  \left( \begin{array}{cc} c_\theta & -s_\theta \\  
                           s_\theta & c_\theta \end{array} \right)
  \left( \begin{array}{c} \bS \\ \bV_3 \end{array} \right) \; ,
  \label{eq:transformation}
\end{eqnarray}
where we have used the notation $c_\theta \equiv \cos\theta$,
$s_\theta \equiv \sin\theta$, and the neutral state mixing angle is
\begin{eqnarray}
\tan 2\theta &=& 
    \frac{2\delta^2}{m_{\bV}^2-m_{\bS}^2} \, .
\end{eqnarray}
The charged squirk masses remain unchanged with mass $m_\pm = M_{\bV}$. 

Due to infracolor confinement, the squirks remain in a bound state.  
The dynamics of the squirk bound states will
occupy much of the later discussion of the paper.
We work in the approximation that 
$\alphaic(m_q) \ll 1$, so that the masses of the bound states 
are dominated by the constituent squirk masses.  
There is a hierarchy of mesons formed from orbital excitations
of the bound squirks.  We will be concerned mainly with just
the $J=L=0$ (with necessarily $S=0$ for squirks) ground states, 
which we write as $\eta_{ij}$ with $ij$ labeling the constituent
squirks.  The charged bound states consist of 
\begin{eqnarray}
\eta_{\pm 0} &\sim& (u_\pm \nzero^\dagger), (\nzero u_\mp^\dagger) \\
\eta_{\pm 1} &\sim& (u_\pm \none^\dagger), (\none u_\mp^\dagger) \; , 
\end{eqnarray}
where the charged meson masses are roughly 
$m_{\eta_{\pm i}} \simeq m_\pm + m_i$.
The neutral states consist of several distinct mass eigenstates
\begin{eqnarray}
\eta_{00} &\sim&  (\nzero \nzero^\dagger) \\
\eta_{+-} &\sim&  (u_+ u_+^\dagger), (u_- u_-^\dagger) \; . \\
\eta_{11} &\sim&  (\none \none^\dagger) 
\end{eqnarray}
with masses that are again roughly 
$m_{\eta_{00}} \simeq  2 m_0$, 
$m_{\eta_{+-}} \simeq  2 m_\pm$, and
$m_{\eta_{11}} \simeq  2 m_1$.
Whether the heavier squirks hadronize before weak decay,
like the $b$-quark of the SM, or decay before hadronization,
like the $t$-quark of the SM, depends on the parameters of
the model.  As we will see, both regimes are relevant to the
model.

%%%%%%%%%%%%%%%%%%%%%%%%%%%%%%%%%%%%%%%%%%%%%%%%%%%%%%%%%%%%%%%%%%%%%%%%%
\section{Squirk Production}
\label{sec:quirkproduction}

Squirks can be produced through color and electroweak 
production.  Conservation of infracolor implies squirks are always 
produced in pairs.  At hadron colliders, squirks are generically 
produced with some kinetic energy for each squirk.
As vividly demonstrated in \cite{Kang:2008ea},
squirks will initially fly apart forming an infracolor 
string between them.  The string will stretch without tearing until the 
energy in the infracolor string matches the total kinetic energy.  
The infracolor scale of interest to us is roughly 
$\LQCD \lsim \Lic \ll m_{\rm squirk}$, where the infracolor strings 
are \emph{microscopic}, and thus the squirks shed their kinetic energy 
and annihilate on timescales short compared with the timing systems 
of the collider detectors.  

Here we consider the several possible 
combinations of squirk pair production in terms of the 
squirk mass eigenstates.

\subsection{$u_\pm \nzero^\dagger + \nzero u_\mp^\dagger$}
\label{sec:upm_nzero}

\begin{figure}[!t]
%\vspace{- 0.5cm}
 \centering
 \includegraphics[width=0.48\textwidth]{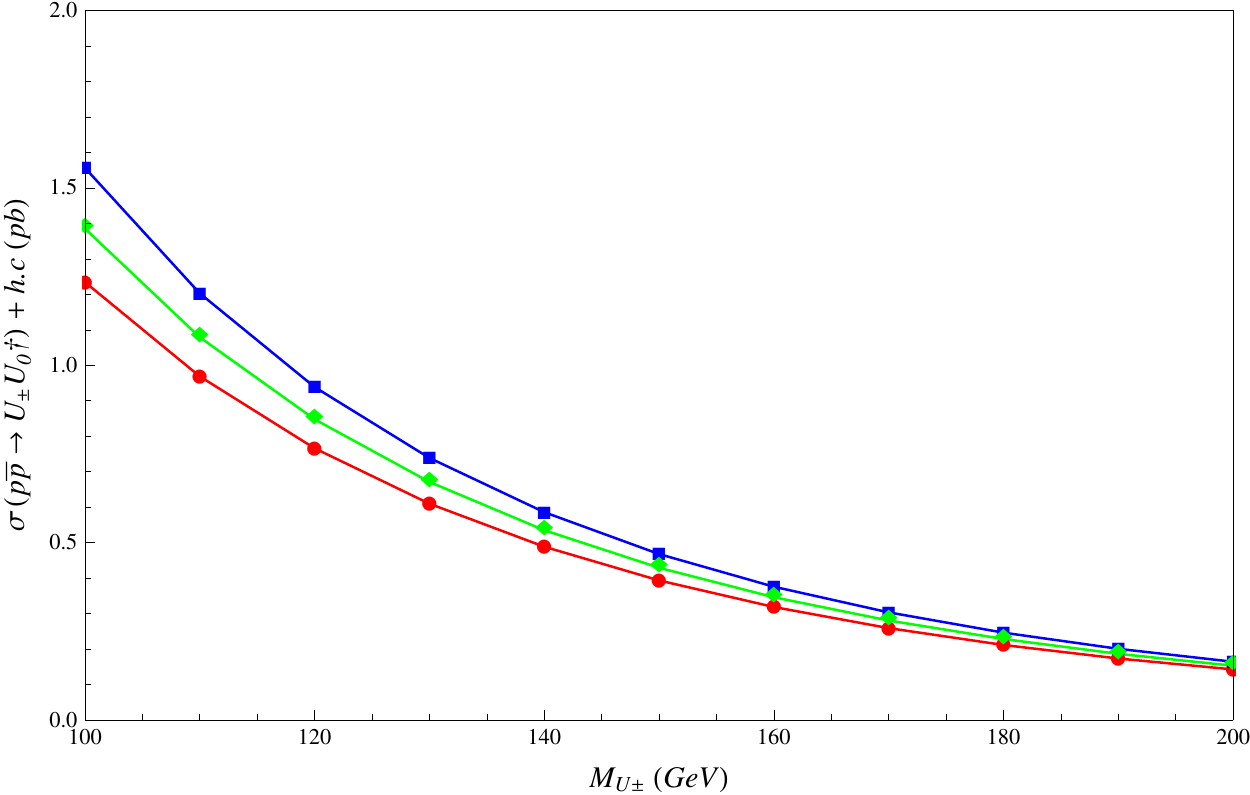}
\caption{Leading order production cross section at Tevatron
$\sigma(u_\pm \nzero^\dagger + \nzero u_\mp^\dagger)$.
The three contours correspond to three choices of 
$m_0 = 75,80,85$~GeV from top to bottom.  
The cross section shown in this figure includes 
the QCD color factor and both electric charges, but \emph{does not} 
include the infracolor (enhancement) factor, nor the mixing angle 
(suppression) associated with the $\bV_3$ component of $\nzero$.}
\label{fig:tevatronweakproduction}
\end{figure}

The production of charged plus neutral squirks proceeds
through a $s$-channel $W^\pm$.  This cross section is enhanced 
by the number of QCD colors and infracolors, while suppressed
by the mixing angle associated with the $\bV_3$ component of $\nzero$.
The cross section at the Tevatron is shown in
Fig.~\ref{fig:tevatronweakproduction}, 
incorporating the QCD color factor enhancement, but without 
the infracolor enhancement as well as without the neutral state 
mixing angle suppression, to remain as general as possible 
at this point.

%%%%%%%%%%%%%%%%%%%%%%%%%%%%%%%%%%%%
%%%%%%%%%%%%%%%%%%%%%%%%%%%%%%%%%%%%

Once squirks are produced, numerous effects must be considered
to obtain a realistic estimate of the signal rates.  
This includes understanding the spin-down process, $\beta$-decay, 
and the various annihilation channels.
In addition, signal efficiencies are non-trivial when 
$m_\pm - m_0 \lsim M_W$ and thus the $W$ decay is off-shell.
This case, which is relevant for our model, implies 
the lepton $p_T$ cuts tend to slice away more signal 
while also changing the dijet kinematics. 
A full estimation of our signal efficiency is not possible 
given the unknowns regarding the quirky spin-down dynamics.
Instead, we have performed estimates of signal efficiency by
constructing a ``stand in'' model, similar to Ref.~\cite{Eichten:2011sh}.
This is discussed in Sec.~\ref{sec:eff}.

\subsection{$\nzero \nzero^\dagger$ and $\none \none^\dagger$}
\label{sec:nzero_nzero}

The electroweak quantum numbers of $\nzero$ and $\none$ 
($T_3 = 0$, $Q = 0$) imply the couplings to $\gamma$ and $Z$ 
exactly vanish, which has important implications.
The first, and perhaps most important, is that $\nzero$ and $\none$
cannot be pair-produced at LEP II at tree-level.\footnote{There 
is a set of one-loop diagrams involving virtual $W$s 
and virtual $u_\pm$ exchange, but this is very small.}
Second, the bound states formed from $\nzero\nzero^\dagger$
and $\none\none^\dagger$ cannot decay into pairs of photons 
or (potentially off-shell) $Z$s.

\begin{figure}[!t]
%\vspace{-0.5cm}
 \centering
 \includegraphics[width=0.48\textwidth]{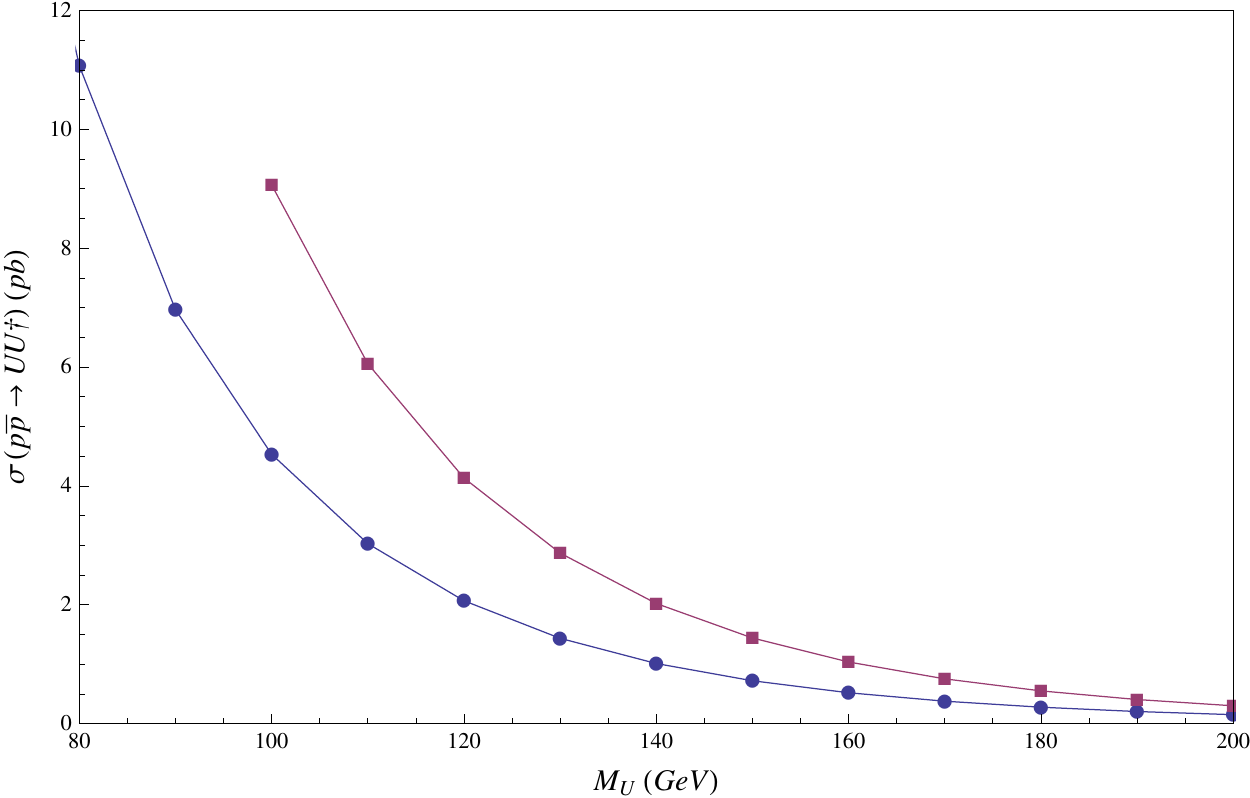}
\caption{Leading order production cross sections at Tevatron
for $\sigma(u_\pm u_\pm^\dagger)$ with (top curve; $M_U \ra m_\pm$) 
and $\sigma(\nzero \nzero^\dagger)$ 
or $\sigma(\none \none^\dagger)$ 
(bottom curve; $M_U \ra m_0$ or $m_1$).
Again, like Fig.~\ref{fig:tevatronweakproduction}, the
cross sections include the QCD color factor and both
electric charges, but do not include the infracolor factor.}
\label{fig:tevatroncoloredproduction}
\end{figure}

The neutral squirks, $\nzero$ and $\none$, can be pair-produced through QCD\@.
The Tevatron production cross section of neutral squirk pairs 
is shown in Fig.~\ref{fig:tevatroncoloredproduction}.
The lightest bound state of squirks must ultimately annihilate 
back into SM particles.
There are several final state topologies that can result:
$gg$ and $g'g'$ (infragluon pair), illustrated in 
Fig.~\ref{fig:tevatroncoloredproduction}, as well as 
$W^+W^-$ (not shown).  

Although our model has negligible LEP II production of squirks,
it is amusing to consider the size of the cross section
for generalized squirks with ($Q, T_3$) arbitrary.
We find that cross section for colored squirks is roughly 
$0.6$~pb for $Q = 0, |T_3| = 1/2$.  The cross section
increases if $Q \not= 0$ is taken.  Since the total hadronic 
cross section at LEP II is measured to within $\pm 0.5$~pb 
to 95\% CL \cite{lepdifermion}, 
even without considering the distinctive kinematics (dijet invariant mass peak),
we see that $80$~GeV squirks transforming under the electroweak group 
are essentially ruled out.  This is the motivation for our model 
employing an electroweak triplet that carries vanishing hypercharge.

%%%%%%%%%%%%%%%%%%%%%%%%%%%%%%%%%%%%
%%%%%%%%%%%%%%%%%%%%%%%%%%%%%%%%%%%%

\subsection{$u_\pm u_\pm^\dagger$}
\label{sec:upm_upm}

Charged squirk pair production, possible $\beta$-decay, spin-down, 
and annihilation is a fascinating but intricate process.  
Charged squirk pairs can be produced through QCD, and so their
production rate at hadron colliders is large.  The cross section at
the Tevatron is shown in Fig.~\ref{fig:tevatroncoloredproduction}. 

The signature of charged squirks depends crucially on the 
various competing timescales of spin-down, $\beta$-decay, and
annihilation.  If both charged squirks were to rapidly $\beta$-decay, 
$u_\pm \ra W^\pm \nzero$,
this leads to a $W^+ W^- \nzero \nzero^\dagger$ signal,
with $\nzero \nzero^\dagger$ annihilating into $gg$ 
(or invisibly to $g'g'$).  
Whether this process is a \emph{feature} or a \emph{constraint} 
depends on several factors, particularly, the mass splitting 
$m_\pm - m_0$ and the associated efficiency of detecting 
the $W$ decay products. 

There is, however, another process that competes with $\beta$-decay,
namely the squirky spin-down and annihilation.  As we will show
in the next few sections, there is a region of parameter space 
where the spin-down and annihilation of $u_\pm u_\pm^\dagger$ occurs 
\emph{faster} than $\beta$-decay, leading to $gg$, $g'g'$, 
$\gamma\gamma$, and $W^+W^-$ signatures.

\subsection{$u_\pm \none^\dagger + \none u_\mp^\dagger$}
\label{sec:upm_none}

The production of the charged plus the \emph{heavier} neutral squirk 
proceeds through a $s$-channel $W^\pm$ just like Sec.~\ref{sec:upm_nzero}.  
This cross section is enhanced 
by the number of QCD colors and infracolors, while suppressed
by the mixing angle of the $\bV_3$ component of $\none$.
While this process has a substantially smaller cross section than
the one with $\nzero$, there are several interesting signatures.  
First, if the $\beta$-decay process $\none \ra W^\pm u_\mp$ occurs quickly, 
then this process contains the same rich phenomenology of 
Sec.~\ref{sec:upm_upm} along with an additional $W^\pm$, which itself 
could be off-shell.  
If instead spin-down occurs rapidly, then a competition is set up
between $\beta$-decay with subsequent annihilation of $\eta_{+-}$,
versus direct annihilation $\eta_{+1} \ra W\gamma$.  
It is also possible that $\beta$-decay of $u_\pm \ra W^\pm \nzero$ 
occurs, forcing $\none \ra W^\pm u_\mp$ followed by $u_\mp \ra W^\mp \nzero$.
This gives a spectacular three-$W$ plus 2-jet signal.

As we will see in the next several sections, the production 
of $\nzero \nzero^\dagger$, $\none \none^\dagger$, and $u_\pm u_\pm^\dagger$ 
are most relevant for the $Wjj$ signal.

%%%%%%%%%%%%%%%%%%%%%%%%%%%%%%%%%%%%
\begin{figure}[!t]
%\vspace{- 0.5cm}
 \centering
 \includegraphics[width=0.48\textwidth]{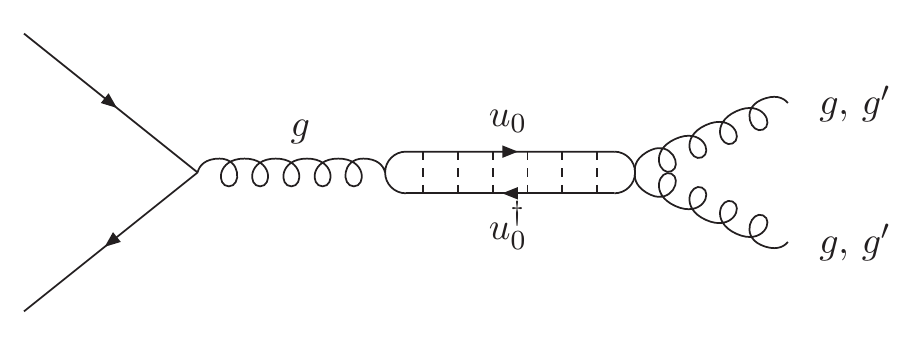}
\caption{One of several diagrams illustrating squirky pair production 
through QCD of the lightest electrically-neutral squirks 
$p\bar{p} \ra \nzero \nzero^\dagger$ and 
annihilation $\nzero \nzero^\dagger \ra gg$.
Radiation (into color gluons, infracolor glueballs, etc.) 
is not shown.}
\label{fig:lightlightquirkyproduction}
\end{figure}
%%%%%%%%%%%%%%%%%%%%%%%%%%%%%%%%%%%%

%%%%%%%%%%%%%%%%%%%%%%%%%%%%%%%%%%%%%%%%%%%%%%%%%%%%%%%%%%%%%%%%%%%%%%%%%
\section{Weak Decay and Spin-down}
\label{sec:spindowndecay}

We now consider the quirky dynamics that occurs after the 
squirk pair is produced.  This includes $\beta$-decay, 
spin-down, and the annihilation of the squirks into SM fields. 
The annihilation cross section itself is a sensitive decreasing function 
of the angular momentum of the squirk system. The angular momentum 
of the system is of order $L\sim1$ in the hard production, 
and then grows as the excited state losses energy through radiation. 
Since every radiated quantum is expected to change the angular momentum 
by $\Delta L \sim \pm 1$, $L$ grows roughly as the square root of the 
number of emitted quanta. As a result, the squirky bound state typically 
loses its excitation energy \emph{before} annihilation~\cite{Kang:2008ea}. 
On the other hand, $\beta$-decay can occur during the energy loss process,
as it is independent of angular momentum. 
Thus, in this section we discuss the competition between 
$\beta$-decay and energy loss, and then in Sec.~\ref{sec:annihilation} 
we calculate the decay rates of squirkonium states formed only after 
energy loss has occurred.

\subsection{$\beta$-decay}
\label{sec:betadecay}

Weak decay of bound squirks or quirks can be readily approximated 
by considering the decay of an individual squirk in isolation.  
This is entirely analogous to heavy quark decay of (heavy)
quarkonia in the SM\@. 
Charged and heavy neutral squirks can $\beta$-decay through
$u_\pm \ra W^\pm \nzero$ and $\none \ra W^\pm u_\mp$.
When the mass splitting among these states 
$\Delta m_i = |m_\pm - m_i|$ is smaller than $M_W$, which is the main
parameter region of interest to us, the rate is given by
\begin{equation}
\Gamma_\beta(u_\pm \ra W^\pm \nzero) = 
\frac{N_f G_F^2 s_\theta^2 R(m_\pm,m_0,m_W)}{48 \pi ^3}
    \label{eq:pmbetadecay}
\end{equation}
where the function $R$ is presented in its full glory in 
Appendix~\ref{sec:formulae}.
The kinematic function $R$ scales approximately as $|\Delta m|^5$,
as expected for a 3-body decay.  
The mixing angle is defined by the transformation given in 
Eq.(\ref{eq:transformation}).
The rate scales as the number of SM flavors and colors 
to decay into, $N_f$, which is $9$ for $m_\tau < |\Delta m| < m_t + m_b$.
The rate for $\none \ra W^\pm u_\mp$ is analogous, 
\begin{equation}
\Gamma_\beta(\none \ra W^\mp u_\pm) = 
\frac{N_f G_F^2 s_\theta^2 R(m_1,m_\pm,m_W)}{24 \pi^3}\; ,
    \label{eq:nonebetadecay}
\end{equation}
where the relative factor of $2$ (compared with the charged squirk case)
accounts for the two possible electric charge combinations.
The inverse decay rate is roughly 
\begin{equation}
t_{\beta,u_\pm} \sim 
    (3 \times 10^{-21} \; {\rm sec}) \times   % 3.2 
    \left(\frac{\mbox{40 GeV}}{\Delta m_0}\right)^5
    \left(\frac{0.717}{s_\theta}\right)^2 \; .
\label{eq:betatimescale}
\end{equation}

\subsection{Spin-down}
\label{sec:spindown}

The ``spin-down'' or energy loss process for squirks transforming
under both QCD and infracolor is an interesting but difficult problem.  
In general, the energy loss can proceed through the radiation of 
a) regular QCD hadrons -- mostly charged and neutral pions, or 
b) infracolor glueballs.\footnote{Photons may also play an 
important role~\cite{Harnik:2008ax} in the case where quirks 
or squirks are charged and uncolored, but this will be 
subdominant to QCD radiation for our model.}
Given the intrinsically non-perturbative nature of this emission,
we will not assert which of these two distinct processes dominates, 
but instead simply make estimates assuming one or the other does, 
and discuss the phenomenology.

One reasonable approach to the mechanism of energy loss is to assume 
that at every squirky oscillation, the two QCD-hadronized squirks
at the end of the infracolor string collide at semi-relativistic 
velocities, emitting a soft hadron, carrying energy roughly 
of order $\LQCD$ or $\Lic$ \cite{Kang:2008ea,Luty-et-al}.
The period of the squirk oscillations is of order 
$\mu/\Lic^2$, where $\mu$ is the reduced mass 
of the two-squirk system. 

Consider first the case where only QCD dominates and the excited 
squirky system loses energy of order $\Delta E \sim \LQCD \sim 1 \; {\rm GeV}$ 
after every oscillation.\footnote{It is not 
unreasonable for the energy loss process to be dominated by 
QCD due to kinematics.  This is because QCD contains pions, which 
are lighter than glueballs, unlike the situation with infracolor.}
The time to lose the 
total excitation energy $E_{ij} = \sqrt{\hat s}-(m_i+m_j)$ 
for ($i,j = 0,1,\pm$) is roughly
\begin{eqnarray}
\label{eq:tloss}
t_{\mathrm{loss}}(E_{ij}) &\sim& 
    \frac{\mu}{\Lic^2}
    \frac{E_{ij}}{\LQCD} \\
&\sim& 
    \left( \frac{\mu}{50 \; {\rm GeV}} \right)
    \left( \frac{1 \; {\rm GeV}}{\Lic} \right)^2
    \left( \frac{E_{ij}}{100 \; {\rm GeV}} \right)
    \left( \frac{1 \; {\rm GeV}}{\LQCD} \right) \nonumber \\
& &{}
    \times (3 \times 10^{-21} \; \mbox{sec}) \; .
    \nonumber 
\end{eqnarray}
For the case $(i,j) = (\pm,0)$, the reduced mass $\mu = 48 \ra 53$~GeV
for $m_\pm \ra 120 \ra 160$~GeV\@.  
Thus, we see that a typical excited squirkonium system settles down 
to its ground state on a timescale roughly comparable to $\beta$-decay,
Eq.~(\ref{eq:betatimescale}), when $|\Delta m| \simeq 40$~GeV\@.
We stress that this mass difference is a rough approximation,
given that we cannot determine the energy loss rate precisely.
In this scenario, every squirky event 
will be accompanied by a large multiplicity of soft pions which will 
contribute to the underlying event. Such new contributions to 
underlying event physics were studied in~\cite{Harnik:2008ax,Luty-et-al}. 

Next, consider the case where the energy is lost also through 
infracolor glueball emission.  This emission proceeds in addition 
to the QCD emission described above, and so the timescale 
in Eq.~(\ref{eq:tloss}) is expected to be an upper bound. 
Infracolor glueballs generically have long lifetimes \cite{Kang:2008ea}, 
and thus will escape the detector as missing energy. 
If infracolor glueball emission does dominate the spin-down process,
then there is additional missing energy in every squirky event.
This affects the $p_T$ distribution and particularly the $\slashchar{E}_T$
distributions of squirk signals, but we will not study this effect 
here (for a different study on the effects of hidden radiation 
see~\cite{Carloni:2010tw}).

\subsection{Survival Probability}
\label{sec:survival}

We can be more precise about our estimate for the fraction of squirk
production that $\beta$-decays before settling into the ground state $\eta$.
This calculation involves convoluting the differential distribution 
$d \sigma_{ij}/d E_{ij}$ with the survival probability,
$P(E_{ij}) = 1 - \exp[-t_{\mathrm{loss}}(E_{ij})/t_{ij}]$ 
to obtain branching fractions:
\begin{equation}
\BR(u_\pm \nzero^\dagger \ra \eta_{\pm 0}) = \frac{1}{\sigma_{\pm 0}}
   \int d E_{\pm 0} \frac{d\sigma_{\pm 0}}{d E_{\pm 0}} P(E_{\pm 0}) 
   \label{eq:survival1}
\end{equation}
\begin{equation}
\BR(u_\pm \nzero^\dagger \ra W^\pm \nzero\nzero^\dagger) = 
    1 - \BR(u_\pm \nzero^\dagger \ra \eta_{\pm 0}) \; , \label{eq:survival2} 
\end{equation}
where $t_{\pm 0} = t_{\beta,u_\pm}$, and 
\begin{equation}
\BR(u_+ {u_+}^\dagger \ra \eta_{+-}) = \frac{1}{\sigma_{+-}}
   \int d E_{+-} \frac{d\sigma_{+-}}{d E_{+-}} P(E_{+-}) \label{eq:survival3}
\end{equation}
\begin{equation}
\BR(u_+ {u_+}^\dagger \ra W^+W^- \nzero \nzero^\dagger) = 
    1 - \BR(u_+ {u_+}^\dagger \ra \eta_{+-}) \; . \label{eq:survival4} 
\end{equation}
where $t_{+-} = t_{\beta,u_\pm}/2$, where the factor of $1/2$ 
accounts for $\beta$-decay of either squirk.  Lastly, 
for completeness, we emphasize that the lightest squirks
cannot $\beta$-decay, and thus lose energy to the ground state 
with unity probability, i.e., 
\begin{equation}
\BR(\nzero\nzero^\dagger \ra \eta_{00}) = 1 \; .
\end{equation}
Here this branching ratio assumes the squirks lose energy 
before they annihilate, which is our working assumption for
this paper \cite{Kang:2008ea,Harnik:2008ax}.
We use these results for our signal estimates in 
Sec.~\ref{sec:horserace}.

\section{Annihilation}
\label{sec:annihilation}

Squirk pairs will eventually settle into a ground state, $\eta$, 
which itself decays through constituent squirk annihilation
analogous to quark annihilation leading to meson decay in QCD.  

In this section we compute the decay rates of the various 
squirkonium states into SM fields and infragluons. 
The formalism for calculating the annihilation rates is similar to that 
of calculating the decay rate of quarkonium in QCD~\cite{Barger:1987xg} 
which has been adapted to quirkonium 
in~\cite{Kang:2008ea,Burdman:2008ek,Cheung:2008ke,FK2011}. 
Moreover, squirkonium has a close relative in supersymmetry,
``stoponium'', whose decay rates have computed 
\cite{Drees:1993uw,Martin:2008sv,Martin:2009dj}, 
and we use to cross-check our own results.
Under the assumption that the squirks settle into the
ground state, $\eta$ ($J^{PC} = 0^{-+}$), 
the decay of $\eta \ra \bar{f} f$ is negligible,
and will not be considered for the remainder of the paper.

The squirkonium decay rate is
\begin{equation}
\Gamma\sim \sigma v_\mathrm{rel} \left|\psi (0)\right|^2
\end{equation}
where $\sigma$ is the annihilation cross section in question, 
$v_\mathrm{rel}$ is the relative velocity among the quirks and 
$\left|\psi(0)\right|^2$ is the squared wave function of the squirkonium 
bound state evaluated at zero squirk separation. The annihilation flux 
$\sigma v_\mathrm{rel}$ is evaluated near threshold, where the two squirks 
are nearly at rest.  Assuming the squirk wave function is Coulombic, 
the squared wave function is
\begin{eqnarray}
\left|\psi (0)\right|^2 = c_\eta\frac{a_0^{-3}}{\pi}
\end{eqnarray}
where $a_0$ is the Bohr radius given by
\begin{eqnarray}
a_0 &=& \left[\left(   C_2(\mathbf{3}) \alphas(a_0^{-1}) 
                      + C_2(\mathbf{N}_{ic}) \alphaic(a_0^{-1}) \right)
         \mu\right]^{-1} .
\end{eqnarray}

Note that there may be significant deviations from the Coulombic result 
as well as other QCD and infracolor effects.  This can be partly
parameterized by a coefficient, $c_\eta$, as discussed in 
Ref.~\cite{Kribs:2009fy}.
In practice, higher order QCD and infracolor effects can 
significantly modify the Coulombic estimate, as can be found by scaling 
the QCD results of \cite{Beneke:2007pj} to include infracolor 
as well as color.  
An additional assumption in our calculations is that we assume the 
$\Nic$ dependence of the wavefunction associated with 
the initial bound state of a pair of squirks is to be treated 
the same way as the $N_c$ dependence of QCD for quarkonia.  
In any case, we do not attempt to model the squirkonium potential
beyond the Coulombic approximation, and thus take $c_\eta = 1$ 
for our estimates of the absolute widths of $\eta$.

Now we consider the several possible annihilation channels for the 
squirky mesons according to the constituent squirk 
states.\footnote{There is also the possibility that quirks 
or squirks could form ``hybrid mesons'' in which the quirk pair is in an infracolor non-singlet state~\cite{hybrid}, leading to
a set of qualitatively different decay processes.  
These include the single emission of glue or infraglue, combined
with another SM gauge boson.  The dynamics of infracolor
non-singlets is very difficult to estimate without a more complete 
picture of the quirky hadronization dynamics, and so we 
do not consider it further.}

\subsection{$\eta_{00}$, $\eta_{11}$}
\label{sec:etazerozero}

All of the neutral mesons can decay into QCD gluons
and infracolor gluons, with decay rates
\begin{eqnarray}
\Gamma(\eta_{ii} \ra gg) &=& 
    \frac{4 \pi \Nic \alphas^2}{3 m_i^2} |\psi(0)|^2 
    \label{eq:etaneutralgg} \\
\Gamma(\eta_{ii} \ra g'g') &=& 
    \frac{3 \pi (\Nic^2 - 1) \alphaic^2}{2 \Nic m_i^2} |\psi(0)|^2 \; ,
    \label{eq:etaneutralgpgp}
\end{eqnarray}
for $i = 0,1$. 
These rates are parton-level approximations without 
QCD or infracolor hadronization.  In the limit $\LQCD,\Lic \ll m_0$, 
the effects of hadronization on the decay widths 
is small \cite{Chackothanks}.  Since $\nni$ does 
not carry electric charge, the tree-level width of 
$\eta_{ii} \ra \gamma\gamma$ vanishes.
There is also potentially a tree-level rate into $W^+ W^-$ through a 
$t/u$-channel $u_\pm$.  This rate is suppressed by kinematics 
(particularly for $\eta_{00} \ra W^+W^-$, since $m_0 \simeq M_W$) 
as well as the weak couplings associated with this channel.  
While potentially interesting for $\eta_{11}$, which can be copiously
produced at LHC, it is subleading compared with the above glue and 
infraglue rates, and we do not consider it further.

The $gg$ channel can dominate, allowing $\nzero\nzero^\dagger$
production at a hadron collider to lead to a dijet bump.  For the
Tevatron, the production rate is far smaller than the QCD background,
leading to neither an observable signal nor a constraint \cite{Aaltonen:2008dn}.
Interestingly, UA2 has comparatively strong bounds on dijet resonances.
The production cross section $\sigma(\nzero\nzero^\dagger)$ at UA2 
($\sqrt s = 630\,\gev$) with $m_0 = 80$~GeV is less than 
$1$~pb per infracolor.
Assuming $\nzero\nzero^\dagger$ quickly spins down and 
annihilates into a dijet resonance of about $160$~GeV, 
the limit we should compare to is the $W'/Z'$ limit at the same mass.
This is $\sigma({\rm dijet}) < 90$~pb for $M_{W'/Z'} = 160$~GeV, 
and so is completely safe. 

The next comparable channel is annihilation into infragluons: $g'g'$.
The signature of the $g'g'$ topology depends on the decay rate 
of the infraglueballs formed from the infragluon emission.
The infraglueball decay rate is generally extremely small, since it proceeds 
through a dimension-8 operator suppressed by squirk masses \cite{Kang:2008ea}.
Infraglueballs are thus expected to escape the detector before they decay. 
In this case, the $g'g'$ signal is itself invisible.  There can also
be a $g g'g'$ process, when the squirks are produced in association 
with an additional initial- or final-state gluon (or quark) jet, 
leading to a monojet signal.

\subsection{$\eta_{\pm 0}$}
\label{sec:etapmzero}

The $\eta_{\pm 0}$ represents an electrically charged squirk plus 
neutral squirk system that has settled into its ground state
before the $u_\pm$ has had a chance to $\beta$-decay.
The dominant annihilation channel is $\eta_{\pm 0} \ra W^\pm \gamma$,
with rate \cite{Burdman:2008ek}
\begin{equation}
\Gamma(\eta_{\pm0} \ra W^\pm\gamma)=
\frac{\pi \alpha \alphaW N_c \Nic s_\theta^2}{(m_0 + m_\pm)^2} f(m_0,m_\pm,m_W)
\left|\psi(0)\right|^2
\label{eq:Wgamma}
\end{equation}
where $\alphaW = \alpha/s_W^2$, and the kinematic function
$f(m_0,m_\pm,m_W)$ has been relegated to the Appendix~\ref{sec:formulae}.
Whether the timescale for this
rate is slower or faster than $\beta$-decay depends on the
parameters in the model.  Small mass splittings tend to suppress
$\beta$-decay, whereas small $\Lic$, which leads to a larger
Bohr radius, tends to suppress the annihilation rate.
There are potentially additional decay channels, including 
$W Z$ and $W h$, but these are much more suppressed by 
phase space, and so we will not calculate their widths.

\subsection{$\eta_{+-}$}
\label{sec:etapmpm}

The neutral meson formed from the two charged squirks can also 
decay into QCD gluons and infracolor gluons, with decay rates
identical to 
Eqs.~(\ref{eq:etaneutralgg},\ref{eq:etaneutralgpgp})
substituting $m_i \ra m_\pm$.
Since the constituent squirks are electrically
charged, $\eta_{+-}$ can also decay into into photons
with a decay rate
\begin{eqnarray}
\Gamma(\eta_{+-} \ra \gamma\gamma) &=& 
    \frac{6 \pi \Nic \alphaem^2}{m_\pm^2} |\psi(0)|^2 \; .
    \label{eq:etaneutralpp}
\end{eqnarray}
This is an interesting annihilation channel, and potentially a 
constraint on our model, given existing bounds on 
$\gamma\gamma$ resonances from collider data.  
We discuss the size of this signal in the next section.

Finally, there is another uniquely squirky annihilation mode,
namely $\eta_{+-} \ra u_0 u_0^\dagger$, through the 
quartic interaction, Eq.~(\ref{eq:quartic}), proportional
to $\lambda_4$.  The rate for this annihilation is given by
\begin{eqnarray}
\Gamma(\eta_{+-} \ra u_0 u_0^\dagger) &=& 
    \frac{9 \pi \Nic^2  
          (\alpha_4 s_\theta^2 + \alpha_{\bV 4} c_\theta^2)^2)}{2 m_\pm^2} 
    \sqrt{1 - \frac{m_0^2}{m_\pm^2}} |\psi(0)|^2 \; .
    \label{eq:wongawonga}
\end{eqnarray}
where $\alpha_{4,\bV 4} \equiv \lambda_{4, \bV 4}/(4 \pi)$.  
This so-called ``double wonga-wonga'' process is unique in that the 
annihilation is from heavy squirks to light squirks without any hard 
SM emission.  In practice, the light squirks will subsequently
spin-down, analogous to the spin-down phase following ordinary 
collider production of squirks, and then the light squirks bind up into
an $\eta_{00}$ and annihilate as discussed in Sec.~\ref{sec:etazerozero}.

%%%%%%%%%%%%%%%%%%%%%%%%%%%%%%%%%%
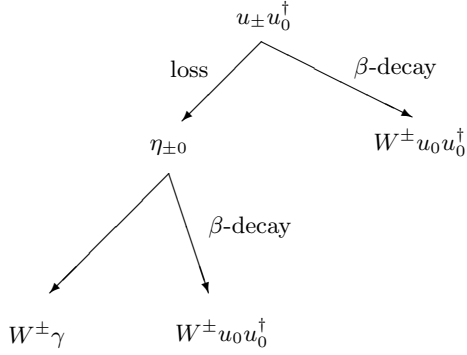
\begin{figure}[!t]
\begin{center}
\begin{picture}(200,140)
\put(101,130){\makebox(0,0){$u_\pm \nzero^\dagger$}}
 \put(100,120){\vector(-1,-1){30}}
 \put( 80,110){\makebox(0,0)[r]{loss}}
 \put(100,120){\vector(+2,-1){57}}
 \put(135,110){\makebox(0,0)[l]{$\beta$-decay}}
\put( 65,80){\makebox(0,0){$\eta_{\pm 0}$}}
\put(160,82){\makebox(0,0){$W^\pm \nzero \nzero^\dagger$}}
 \put(65,70){\vector(-1,-1){45}}
 \put(65,70){\vector( 1,-3){15}}
 \put(15, 9){\makebox(0,0){$W^\pm \gamma$}}
 \put(85,10){\makebox(0,0){$W^\pm \nzero \nzero^\dagger$}}
 \put(80,50){\makebox(0,0)[l]{$\beta$-decay}}
\end{picture}
\end{center}
\caption{The ``tree of life'' for the charged $+$ neutral squirk pair.
While only $u_\pm \nzero^\dagger$ is shown, the same tree of life
also applies to $\nzero u_\pm^\dagger$.}
\label{fig:cntol}
\end{figure}
%%%%%%%%%%%%%%%%%%%%%%%%%%%%%%%%%%

%%%%%%%%%%%%%%%%%%%%%%%%%%%%%%%%%%
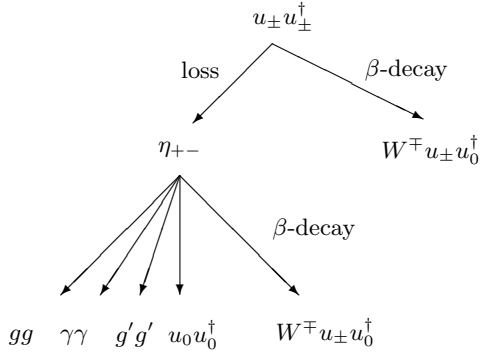
\begin{figure}[!t]
\begin{center}
\begin{picture}(200,140)
 \put(104,130){\makebox(0,0){$u_\pm u_\pm^\dagger$}}
 \put(100,120){\vector(-1,-1){30}}
 \put( 80,110){\makebox(0,0)[r]{loss}}
 \put(100,120){\vector(+2,-1){57}}
 \put(135,110){\makebox(0,0)[l]{$\beta$-decay}}
\put( 65,80){\makebox(0,0){$\eta_{+-}$}}
\put(160,80){\makebox(0,0){$W^\mp u_\pm \nzero^\dagger$}}
 \put( 65,70){\vector(-1,-1){45}}
 \put( 65,70){\vector(-2,-3){30}}
 \put( 65,70){\vector(-1,-3){15}}
 \put( 65,70){\vector( 0,-1){45}}
 \put( 65,70){\vector( 1,-1){45}}
 \put(  5, 9){\makebox(0,0){$gg$}}
 \put( 25, 9){\makebox(0,0){$\gamma\gamma$}}
 \put( 48,10){\makebox(0,0){$g'g'$}}
 \put( 70,10){\makebox(0,0){$\nzero\nzero^\dagger$}}
 \put(120,11){\makebox(0,0){$W^\mp u_\pm \nzero^\dagger$}}
 \put(100,50){\makebox(0,0)[l]{$\beta$-decay}}
\end{picture}
\end{center}
\caption{The ``tree of life'' for the charged squirk pair.
Charged squirk pair tree of life.
Everywhere $u_\pm \nzero^\dagger$ is written is understood to
mean $u_\pm \nzero^\dagger + \nzero u_\mp^\dagger$.}
\label{fig:cctol}
\end{figure}
%%%%%%%%%%%%%%%%%%%%%%%%%%%%%%%%%%

\section{Branching Fractions and Signal Estimates}
\label{sec:horserace}

With the various decay rates in hand we can now compute the 
total cross sections for several final states of interest. 
We will again separate the discussion according to the different 
initial states.

\subsection{$u_\pm \nzero^\dagger + \nzero u_\mp^\dagger$}

The detailed branching fraction ``tree of life'' 
of the quirk pair $u_\pm \nzero^\dagger + \nzero u_\mp^\dagger$
is shown in Fig.~\ref{fig:cntol}.
The first branch splits between $\beta$-decay of $u_\pm$ (right branch)
versus radiative energy loss, leading to the ground state meson 
$\eta_{\pm0}$ (left branch). The relative probabilities 
of these two branches are set by Eqs.~(\ref{eq:survival1}) and 
(\ref{eq:survival2}). 

Following the left branch, in which the ground state
$\eta_{\pm 0}$ is reached, the tree of life then branches 
further, as the meson $\eta_{\pm 0}$ can decay into several different
channels.  Considering all possible $\beta$-decay branchings 
of $u_\pm \nzero^\dagger$ into $W^\pm \nzero\nzero^\dagger$, 
the relevant branching fraction here is
\begin{equation}
\BR(\eta_{\pm 0}\to W^\pm \nzero\nzero^\dagger) = 
    \frac{\Gamma_\beta}{\Gamma_\beta+\Gamma_{W\gamma}}\,
\end{equation}
where the $\Gamma_\beta$ is given by Eq.~(\ref{eq:pmbetadecay}) 
and $\Gamma(W\gamma)$ is given by Eq.~(\ref{eq:Wgamma}).  

Assuming one of the $\beta$-decay branches is taken, the 
tree of life results in the squirk pair $\nzero\nzero^\dagger$,
which after energy loss, ends in the ground state $\eta_{00}$.
The branching fraction of $\eta_{00}$ into visible SM particles
is given by
\begin{equation}
\BR(\eta_{00}\to gg) =  
    \frac{\Gamma(\eta_{00} \to gg)}{
    \Gamma(\eta_{00} \to gg) + \Gamma(\eta_{00} \to g'g')} \, ,
\end{equation}
where the decay rates are given in Sec.~\ref{sec:etapmzero}.

Combining these branching fractions, we obtain the cross section of 
$u_\pm \nzero^\dagger + \nzero u_\mp^\dagger$ into the $Wjj$ final state,
\begin{widetext}
\begin{equation}
\sigma_{\pm 0}(Wjj) =
    \sigma(p\bar p \to u_\pm\nzero^\dagger)
    \left[ \BR(u_\pm\nzero^\dagger \to W^\pm \nzero\nzero^\dagger) + 
           \BR(u_\pm\nzero^\dagger \to \eta_{\pm 0}) 
             \BR(\eta_{\pm 0}\to W^\pm \nzero\nzero^\dagger)
    \right] 
    \BR(\eta_{00} \to gg) \label{eq:sigWjj}
\end{equation}
where the two terms in the square bracket represent the two paths 
to our final state, determined whether $\beta$-decay or energy loss 
occurred first.  The cross section into the $W\gamma$ final state is
\begin{equation}
\sigma_{\pm 0}(W\gamma) = 
   \sigma(p\bar p \to u_\pm\nzero^\dagger)\\
   \left[ \BR(u_\pm\nzero^\dagger \to \eta_{\pm 0}) 
          \BR(\eta_{\pm 0} \to W^\pm \gamma) \right] \, . \label{eq:sigWgamma}
\end{equation}
\end{widetext}
This final state has a bound of about 
$\sigma_{W\gamma} \lesssim 8$-$13$~pb to 95\% CL, 
using constraints from D0 \cite{D0Wgamma} and CDF \cite{CDFWgamma}.
The cross sections given for $\sigma(p\bar{p} \ra u_\pm \nzero^\dagger)$ 
are understood to sum both $u_\pm \nzero^\dagger$ and 
$\nzero u_\mp^\dagger$ contributions.

\subsection{$u_\pm u_\pm^\dagger$}

The tree of life for the $u_\pm u_\pm^\dagger$ is somewhat 
more complicated and is the result of combining 
Figs.~\ref{fig:cctol},\ref{fig:cntol}. The branching begins 
between energy loss and $\beta$-decay (which is now twice as fast 
to account for the decay of either squirk) at the top of Fig.~\ref{fig:cctol}. 
If energy loss is fast, the system will typically get to the 
ground state meson $\eta_{+-}$ (left branch). Here there are 
several interesting decay modes, including $gg$, $g'g'$ (invisible), 
$\gamma\gamma$,\footnote{Note that the final states $W^\pm W^\mp$ 
and $ZZ$ may also be of interest, particularly at LHC, but are 
not calculated here.} and $\nzero\nzero^\dagger$. 
The latter can play an important role in cases where there is 
tension with the $\gamma\gamma$ or pre-tagged top constraints. 
In particular, if $\lambda_4 \sim 2$, the mode Eq.~(\ref{eq:wongawonga}) 
can dominate, 
effectively diminishing the branching fraction into dangerous modes.  
We note however that diphoton and pre-tagged top are safe in 
large regions of our parameter space even in the absence of 
the $\lambda_4$ coupling.

The branching fraction of $\eta_{+-}$ to $\beta$-decay is
\begin{equation}
\BR(\eta_{+-} \to W^\mp u_\pm \nzero^\dagger) =  
    \frac{2 \Gamma_\beta}{2 \Gamma_\beta + 
                          \Gamma_{gg} + 
                          \Gamma_{g'g'} + 
                          \Gamma_{\nzero\nzero^\dagger} + 
                          \Gamma_{\gamma\gamma}} \, ,
\end{equation}
where $u_\pm \nzero^\dagger$ is understood to mean both
$u_\pm \nzero^\dagger$ and $\nzero u_\mp^\dagger$.

We pay close attention to the $WWjj$ final state, which 
is the result of the $\beta$-decay of both squirks and the 
subsequent annihilation of $\eta_{00}$ into dijets. 
Combining branching fractions, we obtain the cross section 
for the ``two-armed lobster'' shown in Fig.~\ref{fig:lobsters} 
into the $WWjj$ final state,
\begin{widetext}
\begin{eqnarray}
\sigma_{+-}(WWjj) &=&
  \sigma(p\bar p \to u_\pm u_\pm^\dagger)
  \left[ \BR(u_\pm u_\pm^\dagger \to W^\mp u_\pm \nzero^\dagger) + 
         \BR(u_\pm u_\pm^\dagger \to \eta_{+-}) 
             \BR(\eta_{+-} \to W^\mp u_\pm u_0^\dagger ) 
    \right] \nonumber \\
& &{} \times 
    \left[ \BR(u_\pm\nzero^\dagger \to W^\pm \nzero\nzero^\dagger) + 
           \BR(u_\pm\nzero^\dagger \to \eta_{\pm 0}) 
             \BR(\eta_{\pm 0}\to W^\pm \nzero\nzero^\dagger) \right]
    \BR(\eta_{00} \to gg) \label{eq:sigWWjj}
\end{eqnarray}
where the two terms in each square bracket represent the four distinct 
paths to our final state, determined whether $\beta$-decay or energy loss 
occurred first in either of Fig.~\ref{fig:cntol} or \ref{fig:cctol}.  
Again, $u_\pm \nzero^\dagger$ is understood to mean 
both $u_\pm \nzero^\dagger$ and $\nzero u_\mp^\dagger$ in the
branching ratios, as appropriate.  
The cross sections of $u_\pm u_\pm^\dagger$ directly into 
various resonance final states are 
\begin{equation}
\sigma_{+-}(ij) = 
    \sigma(p\bar p \to u_\pm u_\pm^\dagger) 
    \left[ \BR(u_\pm u_\pm^\dagger \to \eta_{+-}) 
           \BR(\eta_{+-} \to ij) \right] \, , \label{eq:sigij}
\end{equation}
\end{widetext}
where $ij$ can be any of $gg$, $g'g'$ (invisible), 
$\gamma\gamma$, and $\nzero\nzero^\dagger$. 

The $WWjj$ final state is important for two reasons.  
First, the $WWjj$ signature can ``leak'' into the
$Wjj$ search when the decay products of one $W$ are either lost
or do not pass the CDF analysis cuts for their exclusive analysis.  
In particular, if the mass splitting $m_\pm - m_0$ is not too large
the jets from a hadronically decaying $W$ will be softer, 
and can frequently fall below analysis cuts on jet $E_T$.
The CDF collaboration analysis for $Wjj$ has a 
jet energy requirement of $E_T > 30$ GeV,
suggesting it is much more likely for the jets from a
hadronically decaying off-shell $W$ to \emph{not} pass 
their $E_T$ cuts. 
This means that charged squirk pair production provides 
\emph{an additional production source} of the $Wjj$ signal 
after detector cuts.  
As the mass splitting is increased, a decreasing fraction 
of jets from a hadronically decaying $W$ would \emph{fail} 
their $E_T$ cut, leading to a smaller contribution to the
$Wjj$ signal.

Second, consider the dileptonic process, where both $W$'s decay to
$e$ or $\mu$.  The signature of this process, $l^\pm l^{(')\mp} jj$
plus missing energy, is nearly identical to top quark production
and dileptonic decay, $t\bar{t} \ra W^+ W^- \bar{b} b$, when $b$-tagging 
is \emph{not} required of the jets.  This was analyzed by CDF, 
where their measurements of the fully leptonic ``pre-tagged'' 
top cross section implies an upper limit on this signature of 
about $2$~pb at 95\% CL \cite{CDFdileptonpretag}.

This is likely a reasonable limit for on-shell $W$-pair production
in association with 2 jets.  However, off-shell $W$'s can lead 
to suppression of this signal due to the smaller fraction of events 
with sufficient energy to pass the detector cuts.  
We make some rough estimates of the 
efficiencies of both the $Wjj$ and pre-tagged top analysis in 
the next section. 

Another final state which can potentially constrain the parameter space 
is the decay of the $\eta_{+-}$ meson into $\gamma\gamma$. 
Searches for diphoton resonances were performed by CDF~\cite{Aaltonen:2010cf}, 
D0~\cite{Abazov:2010xh}, and CMS~\cite{Chatrchyan:2011jx} 
in the context of RS graviton searches.
The constraint on the cross section for a diphoton resonance, 
in the mass window of interest to us, $2 m_\pm \simeq 240$-$290$~GeV, 
is of order $10$-$40$~fb.\footnote{A more precise number is 
not straightforward to extract because CDF \cite{Aaltonen:2010cf} 
has a diphoton excess at an invariant mass of 200 GeV, leading to
a weak bound of $100$~fb (as opposed to the expected $30$~fb), 
while D0 \cite{Abazov:2010xh} did not present an exclusion
plot for $\gamma\gamma$ resonance cross section independent of the
$ee$ resonance cross section.}

\section{Benchmarks and Efficiencies}
\label{sec:eff}

We present two benchmarks with the parameters, masses, cross sections,
branching fractions in Table~\ref{table:parameters}.
The two benchmarks represent two qualitatively different
regimes:  Benchmark 1 has rapid spin-down, and then squirky 
$\beta$-decay or annihilation, whereas Benchmark 2 
has rapid $\beta$-decay, and then annihilation. 
Each Benchmark results in qualitatively distinct signal kinematics, 
as we will explain.  
For each Benchmark, we also present our estimates of the relevant 
limits on certain parameters and signal rates.
The input parameters are $\Nic, M_V, M_S, \delta, \lambda_4$.  
We took $\Nic = 3,4$ infracolors for Benchmark 1,2, 
leading to a signal rate into dijets that easily satisfies
the UA2 bound.  We took $M_V, M_S, \delta$ such that
the masses worked out to $m_\pm,m_0 = 120,80$~GeV
and $m_\pm,m_0 = 145,75$~GeV\@.  This allows us to illustrate
the qualitative differences between 
$\Delta m = 40$~GeV versus $\Delta m = 70$~GeV\@.
We chose $M_S$ to be slightly heavier than $M_V$, such that
the contributions to the isospin-violating electroweak correction, 
$\Delta T$, essentially vanish, as described in Appendix~\ref{sec:ewp}.
Finally, we took $\lambda_4$ to be of order one, which ensures
the cross section $\sigma_{+-}(\gamma\gamma)$ is less than
$10$~fb.  Smaller values of $\lambda_4$ imply larger rates,
of order tens of fb, into diphotons.  

%%%%%%%%%%%%%%%%%%%%%%%%%%%
\begin{table}[!t]
\renewcommand{\tabcolsep}{1em}
\begin{tabular}{|c|cc|c|} \hline
                & Bench     & Bench     & Exp't \\ 
                &     1     &     2     & Bound \\ \hline
$\Nic$          & $3$       & $4$       &   -   \\
$\Lic$          & $1.6$     & $6.2$     &   -   \\
$M_V$           & $120$     & $145$     &   -   \\
$M_S$           & $150$     & $250$     &   -   \\
$\delta$        & $106.5$   & $172$     &   -   \\   
$\lambda_4$     & $2$       & $1$       &   -   \\ \hline
$m_0$           & $80$      & $75$      &   -   \\
$m_\pm$         & $120$     & $145$     & $\gsim 100$ \\
$m_1$           & $175$     & $279$     &   -   \\     
$s_\theta$      & $0.82$    & $0.89$    &   -   \\  \hline
$\sigma(\nzero \nzero^\dagger)$ 
                & $33$      & $42$      &   -   \\
$\sigma(u_\pm \nzero^\dagger + \nzero u_\mp^\dagger)$ 
                & $2.5$     & $1.9$     &   -   \\
$\sigma(u_\pm u_\pm^\dagger)$ 
                & $6.2$     & $3.5$     &   -   \\
$BR(\nzero \nzero^\dagger \ra gg)$   
                & $0.51$    & $0.48$    &   -   \\
$BR(\nzero \nzero^\dagger \ra g'g')$ 
                & $0.49$    & $0.52$    &   -   \\
$\sigma_{\rm UA2}(\nzero \nzero^\dagger \ra gg)$ 
                & $0.3$     & $0.6$     & $\lsim 90$ \\
$\sigma_{\pm 0}(Wjj)$
                & $0.72$    & $0.84$    &   -   \\
$\sigma_{+-}(WWjj)$
                & $2.4$     & $2.4$     &   -   \\
$\sigma(\ell^+ \ell^{-(')} jj) \times \mathrm{eff}$
                & $1.6$     & $2.0$     &  $\lsim 2$  \\
$\sigma(W jj) \times \mathrm{eff}$
                & $1.3$-$2.0$ & $1.0$-$1.5$ &  $\lsim 1.9$  \\
$WWjj$/$Wjj_{\rm total}$ 
                & $\sim 85\%$ & $\sim 69\%$ &   -   \\
$\sigma_{+-}(\gamma\gamma)$ 
                & $0.006$   & $0.004$     & $\lsim 0.01$-$0.04$ \\
$\sigma_{\pm 0}(W\gamma)$
                & $1.1$     & $0.2$     & $\lsim 8$-$14$ \\
$\Delta T$      & $0.02$    & $0.01$    & $-0.05 \ra 0.2$ \\ \hline 
$\sigma_{\rm LHC7}(\nzero \nzero^\dagger)$ 
                & $480$     & $430$     &   -  \\
$\sigma_{\rm LHC7}(u_\pm u_\pm^\dagger)$ 
                & $200$     & $130$     &   -   \\ \hline
\end{tabular}
\caption{Benchmark models in parameter space.  All masses in GeV, 
all cross sections are in pb for Tevatron 
(except ``LHC7'' for $\sqrt{s} = 7$~TeV LHC and ``UA2'' for 
$\sqrt{s} = 630$~GeV CERN SppS). 
The cross sections are discussed in Sec.~\ref{sec:quirkproduction},
the branching fractions into various final states discussed in
Sec.~\ref{sec:horserace}, 
the efficiencies are discussed in Sec.~\ref{sec:eff},
and $\Delta T$ calculation is done in Sec.~\ref{sec:ewp}.}
\label{table:parameters}
\end{table}
%%%%%%%%%%%%%%%%%%%%%%%%%%%

The cross sections 
$\sigma(\nzero\nzero^\dagger)$, 
$\sigma(u_\pm \nzero^\dagger + \nzero u_\mp^\dagger)$, and  
$\sigma(u_\pm u_\pm^\dagger)$ can be read off from 
Figs.~\ref{fig:tevatronweakproduction},\ref{fig:tevatroncoloredproduction}.
Weak production is suppressed by couplings, while the colored
production of $\sigma(u_\pm u_\pm^\dagger)$ is enhanced by
couplings but further suppressed by kinematics, leading to
rates at the Tevatron that are roughly comparable.

Each model has $\Lic \ll m_{\rm squirk}$, such that the
infracolor coupling, $\alphaic(m_q)$, is perturbative when 
evaluated at the mass of the squirk.  
The choice $\Lic \sim {\rm few}$~GeV implies 
infraglueballs decay well outside the detector, but 
decay fast enough to not cause cosmological conundrums.
Hence, the $g'g'$ final state leads to no hard SM particles, 
and possibly missing energy (depending on the decay 
kinematics).

The cross sections $\sigma_{\pm 0}(Wjj)$ and $\sigma_{+-}(WWjj)$
correspond to Eqs.~(\ref{eq:sigWjj}) and (\ref{eq:sigWWjj})
respectively.  These cross sections form the starting point
for obtaining the $Wjj$ signal, as well as several additional
signals for which the Tevatron has already placed constraints.

We have then estimated the efficiency to detection for
three signals:  the efficiency for $\sigma_{\pm 0}(Wjj)$
to pass the CDF $Wjj$ analysis cuts \cite{Aaltonen:2011mk}, 
the efficiency for $\sigma_{+-}(WWjj)$ to pass the CDF 
$Wjj$ analysis cuts (where one $W$'s decay products are missed or 
not energetic enough to pass the CDF cuts), and finally, the efficiency 
for $\sigma_{+-}(WWjj)$ to pass the CDF $t\bar{t}$ pre-tag 
analysis cuts \cite{CDFdileptonpretag}, which we
call $\sigma(\ell^+ \ell^{-(')} jj) \times \mathrm{eff}$.

Our estimates are based on a ``stand-in'' model for squirk
production and decay, modeled as intermediate resonances
with masses the same as $\eta_{+-}$, $\eta_{\pm 0}$ and $\eta_{00}$,
allowing for $\beta$-decay into $\eta_{\pm 0}$ which 
in turn can $\beta$-decay into $\eta_{00}$. 
We have implemented both of these models in MadGraph \cite{Alwall:2007st}
with various mass splittings.
The modeling of missing energy cannot be reliably done, 
given the squirky spin-down process that can emit infraglueballs 
which escape the detector as missing energy.  

Nevertheless, we can implement the various lepton, jet,
missing energy and transverse mass cuts on the ``stand in'' 
model to obtain ``stand in'' efficiencies which should
give us a reasonable idea of where the squirk model stands
with respect to the various signals after cuts.
Both benchmarks lead to some excess in the dileptonic pre-tag
$t\bar{t}$ sample ($\sigma(\ell^+ \ell^{-(')} jj)$ plus missing energy),
but within the 95\% CL limit from CDF.
Both benchmarks also lead to between $1$-$2$ pb of $Wjj$ signal,
where the range corresponds to including (not including) 
the missing energy and transverse mass cuts for the lower (upper)
end of the range shown.  
Also shown in Table~\ref{table:parameters} is the ratio 
$WWjj$/$Wjj_{\rm total}$ 
which corresponds to the fraction of
$Wjj$ signal after efficiencies that arise from the
two-armed lobster versus the one-armed lobster.
We conclude that the squirk model
we have presented is capable of generating a $Wjj$ signal
consistent with the CDF excess, so long as the efficiency
of detection of squirks is comparable to our ``stand in''
model for event-level simulation.

While we have presented estimates for a variety of observables above, 
we emphasize that several quantities should be taken as rough
estimates due to the various uncertainties involving quirky dynamics. 
For example, the uncertainty on the timescale of energy loss can strongly
affect the branching fractions between spin-down and $\beta$-decay
as shown in Fig.~\ref{fig:cntol} and \ref{fig:cctol}.
In addition, squirks that rapidly lose energy often reach
the $\eta$ ground states, which have definite masses, and so 
leads to features in kinematic distributions. 
Even when energy loss is rapid, the choice
between branchings involving $\beta$-decay (such as $Wjj$ and $WWjj$) 
versus annihilation channels (such as $W\gamma$ and $\gamma\gamma$) 
is sensitive to the uncertainty in the wavefunction at the origin 
for quirkonium states.

%%%%%%%%%%%%%%%%%%%%%%%%%%%%%%%%%%%%%%%%%%%%%%%%%%%%%%%%%%%%%%%%%%%%
\section{Implications and Discussion}
\label{sec:discussion}

We have demonstrated that the $Wjj$ signal observed by CDF
can be obtained from a model of squirks without violating
existing collider bounds.  There are two qualitative regimes
where the signal arises:

\begin{itemize}

\item[1.] Rapid spin-down, then squirky $\beta$-decay or annihilation.
In this scenario, the excess energy squirks carry, $\sqrt{\hat{s}} - 2 m$, 
is released quickly, allowing the squirk pair production to settle
into squirkonium.  This could happen because $\beta$-decay is 
suppressed, or could happen if the energy loss is rapid, or both.

A competition is set up between $\beta$-decay
of the squirkonium constituent squirks versus direct annihilation
of the squirks.  In this regime, several interesting squirkonium
decay modes with invariant mass resonances are potential 
confirmation signals, including $gg$, $W\gamma$, $\gamma\gamma$, 
$g'g'$ (invisible), and possibly $W^\pm W^\mp$, $W^\pm Z$, $ZZ$,
and signals with a Higgs boson.  

\item[2.] Rapid $\beta$-decay, then spin-down, then annihilation.
In this regime, the only squirkonium state that is reached
is $\eta_{00}$, yielding the famous jet-jet resonance.
This could happen if $\beta$-decay is comparatively fast, 
or could happen if energy loss is comparatively slow, or both.

In this regime, every heavy squirk $\beta$-decay yields a 
(possibly off-shell) $W$, giving many signals with multi-$W$'s
plus a jet-jet resonance.  

\end{itemize}

The two regimes share several interesting signals including
the $Wjj$ final state.  Interestingly, neither regime 
contains an associated $\gamma jj$ or $Z jj$ signal,
which is characteristic of this model.
We emphasize that while squirk production and squirk
$\beta$-decay rates can be calculated perturbatively, 
the ``spin-down'' or energy loss phase, as well as the 
squirkonium decay rates have substantial uncertainties.  
The relative branching fractions of squirkonium can in some cases
be determined, since the decay rate dependence on the 
wave function at the origin drops out.   
But determining which of energy loss or $\beta$-decay 
is more rapid can merely be estimated.

The kinematics of the two regimes to $Wjj$ are qualitatively distinct. 
In particular, if the second term dominates one would expect 
an invariant mass peak for the whole $Wjj$ system, and an edge 
in the transverse mass distribution, while if the second dominates 
such features would be either absent, or less pronounced. 
CDF have presented interesting distributions \cite{CDFWjjwebsite},
but we leave the investigation of such kinematic features for future work.

We have calculated the LHC production cross sections of the
squirks in the parameter ranges relevant to this model.
We find that the electroweak production of charged plus 
neutral squirks is relatively small, typically a few pb.
The colored production of squirks is, not surprising, 
quite large -- of order hundreds of pb!  
The LHC production of $u_\pm u_\pm^\dagger$ seems
particularly important, since depending on which regime we are in, 
it can yield signal rates into the annihilation modes
[regime 1] or the multi-$W$ production [regime 2].

There are several associated signals of our model that
we have not discussed.  
Squirks have nontrivial interactions with the Higgs boson 
through the dimension-4 operator, Eq.~(\ref{eq:dim4}).  
Loops of squirks modify the effective $h g g$ coupling 
as well as $h \gamma\gamma$.  Since $\nzero,\none$ receives 
its mass from both electroweak preserving, Eq.~(\ref{eq:vectormass}) and 
electroweak breaking Eq.~(\ref{eq:dim4}), the contribution
will scale as $\delta^2/(\delta^2 + M^2)$. 

In this paper we have focused mainly on a quirky explanation
for the CDF $Wjj$ excess, but in the process we have illustrated
many interesting signals of quirk or squirk production at
colliders.  With the Tevatron hints of new physics, combined
with the wonderful prospects at LHC, we expect an exciting time
in quirky physics.

\emph{Note added:}  Just before this paper was completed,
D0 reported an analysis of their $Wjj$ data, finding 
``no evidence for anomalous resonant dijet production'' \cite{D0}.
Taken at face value, their analysis sets an upper bound of $1.9$~pb
at 95\% CL for resonant dijet production near the invariant
mass window of the CDF excess, and thus does not rule out a new 
physics explanation
of the CDF excess with a cross section less than this value. 
Given that the D0 analysis was done with lower luminosity 
($4.3$~fb$^{-1}$ \cite{D0} versus $7.3$~fb$^{-1}$ \cite{CDFWjjwebsite}),
without an inclusive (2 or more jets) analysis
(unlike \cite{CDFWjjwebsite}), we remain skeptical of 
drawing negative conclusions until the 
joint task force \cite{jointtaskforce} completes their analysis.

\appendix

%%%%%%%%%%%%%%%%%%%%%%%%%%%%%%%%%%%%%%%%%%%%%%%%%%%%%%%%%%%%%%%%%%%%%%
\section{Electroweak Precision Corrections}
\label{sec:ewp}

The introduction of the Higgs operator at dimension-4, Eq.~(\ref{eq:dim4}),
splits the masses of the fields within the triplet $V$.
This isospin violation leads to modifications to electroweak precision data.
The correction to $S$, which characterizes $B_\mu \leftrightarrow W^0_\mu$
mixing induced by electroweak symmetry breaking, vanishes here
since the squirks transform in electroweak representations 
with zero hypercharge.

The correction to $T$ arises from the mass difference between 
the charged and neutral components of the isotriplets.
This has been calculated for a general scalar multiplet with
arbitrary isospin and hypercharge in Ref.~\cite{Lavoura:1993nq}.
Applying their results to our case, we obtain
\begin{equation}
\Delta T = \frac{N_c \Nic}{16 \pi s_w^2 M_W^2} 
\left[ c^2_\theta f(m_0,m_\pm) - s^2_\theta f(m_1,m_\pm) \right]
\end{equation}
in terms of the self-energy loop function
\begin{eqnarray}
f(m_a,m_b) &=& m_a^2 + m_b^2 - \frac{2 m_a^2 m_b^2}{m_a^2 - m_b^2} 
               \log \frac{m_a^2}{m_b^2}
\end{eqnarray}
As expected for a renormalizable theory, the correction to $T$ 
is finite, and vanishes in the various limits:  $m_0 \ra m_1$ 
(the operator vanishes); $m_V \gg m_S$ (decouple the $V$ scalar); 
and $m_S \gg m_V$ (decouple the $S$ scalar).

Since the model contains negligible additional contributions to 
$\Delta S$, the quantity $\Delta T$ can 
be bounded from fits to $S$ and $T$, e.g.~\cite{Kribs:2007nz}, 
where one can allow $-0.05 \lsim \Delta T \lsim 0.2$ and remain 
within the 95\% CL limits for $m_h = 115$~GeV\@.  
Wide ranges of parameters allow for sizeable splitting between
the charged and neutral squirks (allowing weak decay 
to proceed, c.f.~Sec.\ref{sec:betadecay}), 
while having $T$ easily within the electroweak bounds.

%%%%%%%%%%%%%%%%%%%%%%%%%%%%%%%%%%%%
\begin{figure}[!t]
%\vspace{- 0.5cm}
 \centering
 \includegraphics[width=0.48\textwidth]{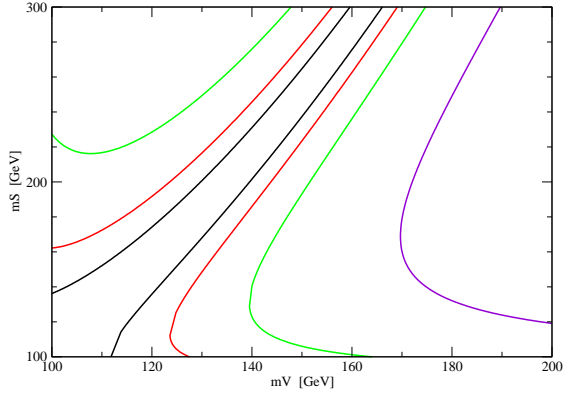}
\caption{Contours of $\Delta T$ \emph{per infracolor}, 
but including QCD color, in the $(m_{\bV},m_{\bS})$ parameter space. 
From left to right the contours are 
$-0.1,-0.05,-0.025,+0.025,0.05,0.1,0.25$.  
The ``funnel'' region has $\Delta T \simeq 0$ due to a cancellation 
between the loops of the light and heavy neutral scalars.}
\label{fig:T}
\end{figure}
%%%%%%%%%%%%%%%%%%%%%%%%%%%%%%%%%%%%

\begin{widetext}
  
\section{Collected Formulae}
\label{sec:formulae}

Here we collect some analytic formulae used in earlier
parts of the paper. The function $f$ used in Eq.~(\ref{eq:sigWgamma}) 
for the weak annihilation of $\eta_{\pm 0}$ into $W\gamma$ is
\begin{equation}
f(m_0,m_\pm,m_W) = \frac{\left[(m_0+m_\pm)^2-m_W^2\right]
 \left[m_W^2 \left(-3 m_0^2+4 m_0 m_\pm+3
   m_\pm^2\right)+m_0 (m_0^2-m_\pm^2) (3
   m_0+m_\pm)\right]}{2 m_0 m_\pm^2 m_W^2 (m_0+m_\pm)}
\end{equation}
The function $R$ used in the formula for the beta decay rate of 
$u_\pm\to u_0 +W^*$ in Eq.~(\ref{eq:pmbetadecay}) is
\begin{eqnarray}
R(m_\pm,m_0,m_W) &=&
 \frac{9 m_W^4 \left(m_\pm^4-m_0^4\right)+2 m_W^2
   \left(m_0^2-m_\pm^2\right)^3  + 6 m_W^6 (m_0^2-m^2_\pm)}{m_\pm^3}
  \nonumber  \\
   & &{} + 6 \frac{m_W^4}{m_\pm^3} 
   \left\{\tilde \Delta^2
   \left(m_0^2+m_\pm^2-m_W^2\right) \left[
   \tan^{-1}\left(\frac{m_0^2-m_\pm^2+m_W^2}{\hat \Delta^2}\right)
   -\tan^{-1}\left(\frac{-m_0^2+m_\pm^2+m_W^2}{\tilde \Delta^2}\right)\right]   
   \right.   \nonumber\\
   & &{} \qquad\qquad - \left.  \left[m_0^4 +m_\pm^4+m_W^4  -2 m_W^2
   \left(m_0^2 +m_\pm^2\right)  \right] \log
   \left(\frac{m_0}{m_\pm}\right)
     \right\}
\end{eqnarray}
where
\begin{equation}
\tilde\Delta^2\equiv \sqrt{(m_0+m_\pm-m_W) (m_0-m_\pm+m_W)
   (-m_0+m_\pm+m_W) (m_0+m_\pm+m_W)}
\end{equation}
and
\begin{equation}
\hat \Delta^2 \equiv \sqrt{(-m_0+m_\pm+m_W) (m_0+m_\pm-m_W) (m_0-m_\pm+m_W)
   (m_0+m_\pm+m_W)}
\end{equation}

\end{widetext}

%%%%%%%%%%%%%%%%%%%%%%%%%%%%%%%%%%%%%%%%%%%%%%%%%%%%%%%%%%%%%%%%%%%%%%%%%%%

\section*{Acknowledgments}

We thank Z.~Chacko, R.~Fok and M.~Strassler for useful conversations. 
GDK was supported by a Ben Lee Fellowship from Fermilab and 
in part by the US Department of Energy under contract number 
DE-FG02-96ER40969.
RH, GDK, AM are supported by Fermilab operated by Fermi Research Alliance, 
LLC under contract number DE-AC02-07CH11359 with the 
US Department of Energy.

%%%%%%%%%%%%%%%%%%%%%%%%%%%%%%%%%%%%%%%%%%%%%%%%%%%%%%%%%%%%%%%%%%%%%%%%%%%


\begin{thebibliography}{99}

%\cite{Aaltonen:2011mk}
\bibitem{Aaltonen:2011mk}
  T.~Aaltonen {\it et al.}  [CDF Collaboration],
  %``Invariant Mass Distribution of Jet Pairs Produced in Association with a $W$
  %boson in $p \bar{p}$ Collisions at $\sqrt{s}= 1.96$ TeV,''
  arXiv:1104.0699 [hep-ex].
  %%CITATION = ARXIV:1104.0699;%%

%\cite{CDFWjjwebsite}
\bibitem{CDFWjjwebsite}
  A.~Annovi, P.~Catastini, V.~Cavaliere, and L. Ristori,
  \url{http://www-cdf.fnal.gov/physics/ewk/2011/wjj/7_3.html}.

%\cite{Eichten:2011sh}
\bibitem{Eichten:2011sh}
  E.~J.~Eichten, K.~Lane and A.~Martin,
  %``Technicolor at the Tevatron,''
  arXiv:1104.0976 [hep-ph].
  %%CITATION = ARXIV:1104.0976;%%

\bibitem{varyinginterest}
%\cite{Buckley:2011vc}
%\bibitem{Buckley:2011vc}
  M.~R.~Buckley, D.~Hooper, J.~Kopp, E.~Neil,
  %``Light Z' Bosons at the Tevatron,''
  [arXiv:1103.6035 [hep-ph]];
%\cite{Yu:2011cw}
%\bibitem{Yu:2011cw}
  F.~Yu,
  %``A Z' Model for the CDF Dijet Anomaly,''
  arXiv:1104.0243 [hep-ph];
  %%CITATION = ARXIV:1104.0243;%%
%\cite{Kilic:2011sr}
%\bibitem{Kilic:2011sr}
  C.~Kilic and S.~Thomas,
  %``Signatures of Resonant Super-Partner Production with Charged-Current
  %Decays,''
  arXiv:1104.1002 [hep-ph];
  %%CITATION = ARXIV:1104.1002;%%
%\cite{Cheung:2011zt}
%\bibitem{Cheung:2011zt}
  K.~Cheung and J.~Song,
  %``Tevatron Wjj Anomaly and the baryonic $Z'$ solution,''
  arXiv:1104.1375 [hep-ph];
  %%CITATION = ARXIV:1104.1375;%%
%\cite{AguilarSaavedra:2011zy}
%\bibitem{AguilarSaavedra:2011zy}
  J.~A.~Aguilar-Saavedra and M.~Perez-Victoria,
  %``No like-sign tops at Tevatron: Constraints on extended models and
  %implications for the t tbar asymmetry,''
  arXiv:1104.1385 [hep-ph];
  %%CITATION = ARXIV:1104.1385;%%
%\cite{He:2011ss}
%\bibitem{He:2011ss}
  X.~G.~He and B.~Q.~Ma,
  %``The CDF dijet excess from intrinsic quarks,''
  arXiv:1104.1894 [hep-ph];
  %%CITATION = ARXIV:1104.1894;%%
%\cite{Wang:2011ta}
%\bibitem{Wang:2011ta}
  X.~P.~Wang, Y.~K.~Wang, B.~Xiao, J.~Xu and S.~h.~Zhu,
  %``New Color-Octet Vector Boson Revisit,''
  arXiv:1104.1917 [hep-ph];
  %%CITATION = ARXIV:1104.1917;%%
%\cite{Sato:2011ui}
%\bibitem{Sato:2011ui}
  R.~Sato, S.~Shirai and K.~Yonekura,
  %``A Possible Interpretation of CDF Dijet Mass Anomaly and its Realization in
  %Supersymmetry,''
  arXiv:1104.2014 [hep-ph];
  %%CITATION = ARXIV:1104.2014;%%
%\cite{Nelson:2011us}
%\bibitem{Nelson:2011us}
  A.~E.~Nelson, T.~Okui and T.~S.~Roy,
  %``A unified, flavor symmetric explanation for the t-tbar asymmetry and Wjj
  %excess at CDF,''
  arXiv:1104.2030 [hep-ph];
  %%CITATION = ARXIV:1104.2030;%%
%\cite{Anchordoqui:2011ag}
%\bibitem{Anchordoqui:2011ag}
  L.~A.~Anchordoqui, H.~Goldberg, X.~Huang, D.~Lust and T.~R.~Taylor,
  %``Stringy origin of Tevatron Wjj anomaly,''
  arXiv:1104.2302 [hep-ph];
  %%CITATION = ARXIV:1104.2302;%%
%\cite{Dobrescu:2011px}
%\bibitem{Dobrescu:2011px}
  B.~A.~Dobrescu and G.~Z.~Krnjaic,
  %``Weak-triplet, color-octet scalars and the CDF dijet excess,''
  arXiv:1104.2893 [hep-ph];
  %%CITATION = ARXIV:1104.2893;%%
%\cite{Zhu:2011ww}
%\bibitem{Zhu:2011ww}
  G.~Zhu,
  %``B physics constraints on a flavor symmetric scalar model to account for the
  %ttbar asymmetry and Wjj excess at CDF,''
  arXiv:1104.3227 [hep-ph];
  %%CITATION = ARXIV:1104.3227;%%
%\cite{Ko:2011ns}
%\bibitem{Ko:2011ns}
  P.~Ko, Y.~Omura and C.~Yu,
  %``Dijet resonance from leptophobic Z' and light baryonic cold dark matter,''
  arXiv:1104.4066 [hep-ph];
  %%CITATION = ARXIV:1104.4066;%%
%\cite{Plehn:2011nx}
%\bibitem{Plehn:2011nx}
  T.~Plehn and M.~Takeuchi,
  %``W+Jets at CDF: Evidence for Top Quarks,''
  arXiv:1104.4087 [hep-ph];
  %%CITATION = ARXIV:1104.4087;%%
%\cite{Fox:2011qd}
%\bibitem{Fox:2011qd}
  P.~J.~Fox, J.~Liu, D.~Tucker-Smith and N.~Weiner,
  %``An Effective Z',''
  arXiv:1104.4127 [hep-ph];
  %%CITATION = ARXIV:1104.4127;%%
%\cite{Jung:2011ue}
%\bibitem{Jung:2011ue}
  D.~W.~Jung, P.~Ko and J.~S.~Lee,
  %``A Possible Common Origin of the Top Forward-backward Asymmetry and the CDF
  %Dijet Resonance,''
  arXiv:1104.4443 [hep-ph];
  %%CITATION = ARXIV:1104.4443;%%
%\cite{Chang:2011wj}
%\bibitem{Chang:2011wj}
  S.~Chang, K.~Y.~Lee and J.~Song,
  %``The CDF dijet excess and Z'_{cs} coupled to the second generation quarks,''
  arXiv:1104.4560 [hep-ph];
  %%CITATION = ARXIV:1104.4560;%%
%\cite{Nielsen:2011wz}
%\bibitem{Nielsen:2011wz}
  H.~B.~Nielsen,
  %``The New Dijet Particle in the Tevatron IS the Higgs,''
  arXiv:1104.4642 [hep-ph];
  %%CITATION = ARXIV:1104.4642;%%
%\cite{Bhattacherjee:2011yh}
%\bibitem{Bhattacherjee:2011yh}
  B.~Bhattacherjee and S.~Raychaudhuri,
  %``Tevatron Signal for an Unmixed Radion,''
  arXiv:1104.4749 [hep-ph];
  %%CITATION = ARXIV:1104.4749;%%
%\cite{Cao:2011yt}
%\bibitem{Cao:2011yt}
  Q.~H.~Cao, M.~Carena, S.~Gori, A.~Menon, P.~Schwaller, C.~E.~M.~Wagner and L.~T.~M.~Wang,
  %``W plus two jets from a quasi-inert Higgs doublet,''
  arXiv:1104.4776 [hep-ph];
  %%CITATION = ARXIV:1104.4776;%%
%\cite{Babu:2011yw}
%\bibitem{Babu:2011yw}
  K.~S.~Babu, M.~Frank and S.~K.~Rai,
  %``Top quark asymmetry and Wjj excess at CDF from gauged flavor symmetry,''
  arXiv:1104.4782 [hep-ph];
  %%CITATION = ARXIV:1104.4782;%%
%\cite{Dutta:2011kg}
%\bibitem{Dutta:2011kg}
  B.~Dutta, S.~Khalil, Y.~Mimura and Q.~Shafi,
  %``Dimuon CP Asymmetry in B Decays and Wjj Excess in Two Higgs Doublet
  %Models,''
  arXiv:1104.5209 [hep-ph];
  %%CITATION = ARXIV:1104.5209
%\cite{Huang:2011ph}
%\bibitem{Huang:2011ph}
  X.~Huang,
  %``Anomaly Puzzle, Curved-Spacetime Spinor Hamiltonian, and String Phenomenology,''
  [arXiv:1104.5389 [hep-ph]];
%\cite{Kim:2011xv}
%\bibitem{Kim:2011xv}
  J.~E.~Kim, S.~Shin,
  %``Z' from SU(6)$\times$SU(2)_h GUT, Wjj anomaly and Higgs boson mass bound,''
  [arXiv:1104.5500 [hep-ph]];
%\cite{Carpenter:2011yj}
%\bibitem{Carpenter:2011yj}
  L.~M.~Carpenter, S.~Mantry,
  %``Color-Octet, Electroweak-Doublet Scalars and the CDF Dijet Anomaly,''
  [arXiv:1104.5528 [hep-ph]];
%\cite{Segre:2011tn}
%\bibitem{Segre:2011tn}
  G.~Segre, B.~Kayser,
  %``A Scalar Doublet at the Tevatron?,''
  [arXiv:1105.1808 [hep-ph]];
%\cite{Enkhbat:2011qz}
%\bibitem{Enkhbat:2011qz}
  T.~Enkhbat, X.~-G.~He, Y.~Mimura, H.~Yokoya,
  %``Colored Scalars And The CDF $W+$dijet Excess,''
  [arXiv:1105.2699 [hep-ph]];
%\cite{Chen:2011wp}
%\bibitem{Chen:2011wp}
  C.~-H.~Chen, C.~-W.~Chiang, T.~Nomura, Y.~Fusheng,
  %``A light charged Higgs boson in two-Higgs doublet model for CDF $Wjj$ anomaly,''
  [arXiv:1105.2870 [hep-ph]];
%\cite{Bettoni:2011nd}
%\bibitem{Bettoni:2011nd}
  D.~Bettoni, P.~Dalpiaz, P.~F.~Dalpiaz, M.~Fiorini, I.~Masina and G.~Stancari,
  %``Experimental proposal to study the excess at $M_{jj}=150$~GeV presented by
  %CDF at Fermilab,''
  arXiv:1105.3661 [hep-ex];
  %%CITATION = ARXIV:1105.3661;%%
%\cite{Liu:2011di}
%\bibitem{Liu:2011di}
  Z.~Liu, P.~Nath and G.~Peim,
  %``An Explanation of the CDF Dijet Anomaly within a $U(1)_X$ Stueckelberg
  %Extension,''
  arXiv:1105.4371 [hep-ph];
  %%CITATION = ARXIV:1105.4371;%%
%\cite{Hektor:2011ms}
%\bibitem{Hektor:2011ms}
  A.~Hektor, G.~Hutsi, M.~Kadastik, K.~Kannike, M.~Raidal and D.~M.~Straub,
  %``Direct detection and CMB constraints on light DM scenario of top quark
  %asymmetry and dijet excess at Tevatron,''
  arXiv:1105.5644 [hep-ph];
  %%CITATION = ARXIV:1105.5644;%%
%\cite{Hewett:2011nb}
%\bibitem{Hewett:2011nb}
  J.~L.~Hewett and T.~G.~Rizzo,
  %``Dissecting the Wjj Anomaly: Diagnostic Tests of a Leptophobic Z',''
  arXiv:1106.0294 [hep-ph];
  %%CITATION = ARXIV:1106.0294;%%
%\cite{Fan:2011vw}
%\bibitem{Fan:2011vw}
  J.~Fan, D.~Krohn, P.~Langacker and I.~Yavin,
  %``A Higgsophilic s-channel Z' and the CDF W+2J Anomaly,''
  arXiv:1106.1682 [hep-ph].
  %%CITATION = ARXIV:1106.1682;%%

%\cite{Campbell:2011gp}
\bibitem{Campbell:2011gp}
  J.~M.~Campbell, A.~Martin and C.~Williams,
  %``NLO predictions for a lepton, missing transverse momentum and dijets at the
  %Tevatron,''
  arXiv:1105.4594 [hep-ph].
  %%CITATION = ARXIV:1105.4594;%%

%\cite{Okun:1980kw}
\bibitem{old-quirks}
  L.~B.~Okun,
  %``Thetons,''
  JETP Lett.\  {\bf 31}, 144 (1980)
  [Pisma Zh.\ Eksp.\ Teor.\ Fiz.\  {\bf 31}, 156 (1979)];
  %%CITATION = ZFPRA,31,156;%% 
%\cite{Okun:1980mu}
%\bibitem{Okun:1980mu}
  L.~B.~Okun,
  %``Theta Particles,''
  Nucl.\ Phys.\  B {\bf 173}, 1 (1980);
  %%CITATION = NUPHA,B173,1;%% 
%\cite{Bjorken:1979hv}
%\bibitem{bj}
  J.~D.~Bjorken, (1979),
  %``Elements Of Quantum Chromodynamics,''
SLAC-PUB-2372;  
  %\cite{Gupta:1981ve}
%\bibitem{quinn}
  S.~Gupta and H.~R.~Quinn,
  %``Heavy Quarks And Perturbative QCD Calculations,''
  Phys.\ Rev.\  D {\bf 25}, 838 (1982).
  %%CITATION = PHRVA,D25,838;%%

%\cite{Kang:2008ea}
\bibitem{Kang:2008ea}
  J.~Kang and M.~A.~Luty,
  %``Macroscopic Strings and 'Quirks' at Colliders,''
  JHEP {\bf 0911}, 065 (2009)
  [arXiv:0805.4642 [hep-ph]].
  %%CITATION = JHEPA,0911,065;%%

%\cite{Burdman:2006tz}
\bibitem{Burdman:2006tz}
  G.~Burdman, Z.~Chacko, H.~S.~Goh and R.~Harnik,
  %``Folded supersymmetry and the LEP paradox,''
  JHEP {\bf 0702}, 009 (2007)
  [arXiv:hep-ph/0609152].
  %%CITATION = JHEPA,0702,009;%%

%\cite{Burdman:2008ek}
\bibitem{Burdman:2008ek}
  G.~Burdman, Z.~Chacko, H.~S.~Goh, R.~Harnik and C.~A.~Krenke,
  %``The Quirky Collider Signals of Folded Supersymmetry,''
  Phys.\ Rev.\  D {\bf 78}, 075028 (2008)
  [arXiv:0805.4667 [hep-ph]].
  %%CITATION = PHRVA,D78,075028;%%

%\cite{Cheung:2008ke}
\bibitem{Cheung:2008ke}
  K.~Cheung, W.~Y.~Keung and T.~C.~Yuan,
  %``Phenomenology of iquarkonium,''
  Nucl.\ Phys.\  B {\bf 811}, 274 (2009)
  [arXiv:0810.1524 [hep-ph]].
  %%CITATION = NUPHA,B811,274;%%

%\cite{Harnik:2008ax}
\bibitem{Harnik:2008ax}
  R.~Harnik and T.~Wizansky,
  %``Signals of New Physics in the Underlying Event,''
  Phys.\ Rev.\  D {\bf 80}, 075015 (2009)
  [arXiv:0810.3948 [hep-ph]].
  %%CITATION = PHRVA,D80,075015;%%

%\cite{Cai:2008au}
\bibitem{Cai:2008au}
  H.~Cai, H.~C.~Cheng and J.~Terning,
  %``A Quirky Little Higgs Model,''
  JHEP {\bf 0905}, 045 (2009)
  [arXiv:0812.0843 [hep-ph]].
  %%CITATION = JHEPA,0905,045;%%

%\cite{Kilic:2009mi}
\bibitem{Kilic:2009mi}
  C.~Kilic, T.~Okui and R.~Sundrum,
  %``Vectorlike Confinement at the LHC,''
  JHEP {\bf 1002}, 018 (2010)
  [arXiv:0906.0577 [hep-ph]].
  %%CITATION = JHEPA,1002,018;%%

%\cite{Chang:2009sv}
\bibitem{Chang:2009sv}
  S.~Chang and M.~A.~Luty,
  %``Displaced Dark Matter at Colliders,''
  arXiv:0906.5013 [hep-ph].
  %%CITATION = ARXIV:0906.5013;%%

%\cite{Nussinov:2009hc}
\bibitem{Nussinov:2009hc}
  S.~Nussinov, C.~Jacoby,
  %``Some Comments on the `Quirks' Scenario,''
  [arXiv:0907.4932 [hep-ph]].
  %%CITATION = ARXIV:0907.4932;%%

%\cite{Kribs:2009fy}
\bibitem{Kribs:2009fy}
  G.~D.~Kribs, T.~S.~Roy, J.~Terning and K.~M.~Zurek,
  %``Quirky Composite Dark Matter,''
  Phys.\ Rev.\  D {\bf 81}, 095001 (2010)
  [arXiv:0909.2034 [hep-ph]].
  %%CITATION = PHRVA,D81,095001;%%

%\cite{Kilic:2010et}
\bibitem{Kilic:2010et}
  C.~Kilic and T.~Okui,
  %``The LHC Phenomenology of Vectorlike Confinement,''
  JHEP {\bf 1004}, 128 (2010)
  [arXiv:1001.4526 [hep-ph]].
  %%CITATION = JHEPA,1004,128;%%

%\cite{Carloni:2010tw}
\bibitem{Carloni:2010tw}
  L.~Carloni and T.~Sjostrand,
  %``Visible Effects of Invisible Hidden Valley Radiation,''
  JHEP {\bf 1009}, 105 (2010)
  [arXiv:1006.2911 [hep-ph]].
  %%CITATION = JHEPA,1009,105;%%

%\cite{Martin:2010kk}
\bibitem{Martin:2010kk}
  S.~P.~Martin,
  %``Quirks in supersymmetry with gauge coupling unification,''
  Phys.\ Rev.\  D {\bf 83}, 035019 (2011)
  [arXiv:1012.2072 [hep-ph]].
  %%CITATION = PHRVA,D83,035019;%%

%\cite{FK2011}
\bibitem{FK2011}
  R.~Fok, G.D.~Kribs, to appear.
  
%\cite{Luty-et-al}
\bibitem{Luty-et-al}
  R.~Harnik, G.~Y.~Huang, M.~Luty, S.~Mrenna, in progress.

%\cite{Strassler:2006im}
\bibitem{Strassler:2006im}
  M.~J.~Strassler and K.~M.~Zurek,
  %``Echoes of a hidden valley at hadron colliders,''
  Phys.\ Lett.\  B {\bf 651}, 374 (2007)
  [arXiv:hep-ph/0604261].
  %%CITATION = PHLTA,B651,374;%%

%\cite{Han:2007ae}
\bibitem{Han:2007ae}
  T.~Han, Z.~Si, K.~M.~Zurek and M.~J.~Strassler,
  %``Phenomenology of hidden valleys at hadron colliders,''
  JHEP {\bf 0807}, 008 (2008)
  [arXiv:0712.2041 [hep-ph]].
  %%CITATION = JHEPA,0807,008;%%

%\cite{Strassler:2008fv}
\bibitem{Strassler:2008fv}
  M.~J.~Strassler,
  %``On the Phenomenology of Hidden Valleys with Heavy Flavor,''
  arXiv:0806.2385 [hep-ph].
  %%CITATION = ARXIV:0806.2385;%%

%\cite{Juge:1997nc}
\bibitem{hybrid}
For example, see  K.~J.~Juge, J.~Kuti, C.~J.~Morningstar,
  %``Gluon excitations of the static quark potential and the hybrid quarkonium spectrum,''
  Nucl.\ Phys.\ Proc.\ Suppl.\  {\bf 63}, 326-331 (1998).
  [hep-lat/9709131].

%\cite{Abazov:2010yb}
\bibitem{Abazov:2010yb}
  V.~M.~Abazov {\it et al.}  [D0 Collaboration],
  %``Search for New Fermions ('Quirks') at the Fermilab Tevatron Collider,''
  Phys.\ Rev.\ Lett.\  {\bf 105}, 211803 (2010)
  [arXiv:1008.3547 [hep-ex]].
  %%CITATION = PRLTA,105,211803;%%

%\cite{Aaltonen:2008dn}
\bibitem{Aaltonen:2008dn}
  T.~Aaltonen {\it et al.}  [CDF Collaboration],
  %``Search for new particles decaying into dijets in proton-antiproton
  %collisions at s**(1/2) = 1.96-TeV,''
  Phys.\ Rev.\  D {\bf 79}, 112002 (2009)
  [arXiv:0812.4036 [hep-ex]].
  %%CITATION = PHRVA,D79,112002;%%

%\cite{Alitti:1993pn}
\bibitem{Alitti:1993pn}
  J.~Alitti {\it et al.}  [UA2 Collaboration],
  %``A Search for new intermediate vector mesons and excited quarks decaying to
  %two jets at the CERN $\bar{p} p$ collider,''
  Nucl.\ Phys.\  B {\bf 400}, 3 (1993).
  %%CITATION = NUPHA,B400,3;%%

%\cite{lepdifermion}
\bibitem{lepdifermion}
  %``LEP II Difermion Results for Summer 2002''
  LEPEWWG $f\bar{f}$ Subgroup, LEP2FF/02-03,
  \url{http://lepewwg.web.cern.ch/LEPEWWG/lep2/summer2002/summer2002.ps}.

%\cite{Barger:1987xg}
\bibitem{Barger:1987xg}
  V.~D.~Barger, E.~W.~N.~Glover, K.~Hikasa, W.~Y.~Keung, M.~G.~Olsson, C.~J.~.~Suchyta and X.~R.~Tata,
  %``SUPERHEAVY QUARKONIUM PRODUCTION AND DECAYS: A NEW HIGGS SIGNAL,''
  Phys.\ Rev.\  D {\bf 35}, 3366 (1987)
  [Erratum-ibid.\  D {\bf 38}, 1632 (1988)]
  [Phys.\ Rev.\  D {\bf 38}, 1632 (1988)].
  %%CITATION = PHRVA,D38,1632;%%

%\cite{Drees:1993uw}
\bibitem{Drees:1993uw}
  M.~Drees and M.~M.~Nojiri,
  %``Production and decay of scalar stoponium bound states,''
  Phys.\ Rev.\  D {\bf 49}, 4595 (1994)
  [arXiv:hep-ph/9312213].
  %%CITATION = PHRVA,D49,4595;%%

%\cite{Martin:2008sv}
\bibitem{Martin:2008sv}
  S.~P.~Martin,
  %``Diphoton decays of stoponium at the Large Hadron Collider,''
  Phys.\ Rev.\  D {\bf 77}, 075002 (2008)
  [arXiv:0801.0237 [hep-ph]].
  %%CITATION = PHRVA,D77,075002;%%

%\cite{Martin:2009dj}
\bibitem{Martin:2009dj}
  S.~P.~Martin and J.~E.~Younkin,
  %``Radiative corrections to stoponium annihilation decays,''
  Phys.\ Rev.\  D {\bf 80}, 035026 (2009)
  [arXiv:0901.4318 [hep-ph]].
  %%CITATION = PHRVA,D80,035026;%%

%\cite{Beneke:2007pj}
\bibitem{Beneke:2007pj}
  M.~Beneke, Y.~Kiyo, A.~A.~Penin,
  %``Ultrasoft contribution to quarkonium production and annihilation,''
  Phys.\ Lett.\  {\bf B653}, 53-59 (2007).
  [arXiv:0706.2733 [hep-ph]];
  %%CITATION = PHLTA,B653,53;%%
  %\cite{Beneke:2007gj}
%\bibitem{Beneke:2007gj}
  M.~Beneke, Y.~Kiyo, K.~Schuller,
  %``Third-order non-Coulomb correction to the S-wave quarkonium wave functions at the origin,''
  Phys.\ Lett.\  {\bf B658}, 222-229 (2008).
  [arXiv:0705.4518 [hep-ph]].
  %%CITATION = PHLTA,B658,222;%%

\bibitem{Chackothanks}
  We thank Z.~Chacko for emphasizing this point to the more
  stubborn author.  

%\cite{D0Wgamma}
\bibitem{D0Wgamma}
  D0 Collaboration, D0 Note 6172-CONF, 
  \url{http://www-d0.fnal.gov/Run2Physics/WWW/results/prelim/EW/E36/E36.pdf}.

%\cite{CDFWgamma}
\bibitem{CDFWgamma}
  B.~Heinemann and A.~Nagano [CDF Collaboration],
  \url{http://www-cdf.fnal.gov/physics/ewk/2007/wgzg/}

%\cite{CDFdileptonpretag}
\bibitem{CDFdileptonpretag}
  CDF Collaboration, Conf.\ Note 10163, 
  \url{http://www-cdf.fnal.gov/physics/new/top/2010/xsection/ttbar_dil_xsec_5invfb/cdfpubnote.pdf}.

%\cite{Aaltonen:2010cf}
\bibitem{Aaltonen:2010cf}
  T.~Aaltonen {\it et al.}  [CDF Collaboration],
  %``Search for Randall-Sundrum Gravitons in the Diphoton Channel at CDF,''
  Phys.\ Rev.\  D {\bf 83}, 011102 (2011)
  [arXiv:1012.2795 [hep-ex]].
  %%CITATION = PHRVA,D83,011102;%%

%\cite{Abazov:2010xh}
\bibitem{Abazov:2010xh}
  V.~M.~Abazov {\it et al.}  [The D0 Collaboration],
  %``Search for Randall-Sundrum gravitons in the dielectron and diphoton final
  %states with 5.4 fb-1 of data from ppbar collisions at sqrt(s)=1.96 TeV,''
  Phys.\ Rev.\ Lett.\  {\bf 104}, 241802 (2010)
  [arXiv:1004.1826 [hep-ex]].
  %%CITATION = PRLTA,104,241802;%%

%\cite{Chatrchyan:2011jx}
\bibitem{Chatrchyan:2011jx}
  S.~Chatrchyan {\it et al.}  [CMS Collaboration],
  %``Search for Large Extra Dimensions in the Diphoton Final State at the Large
  %Hadron Collider,''
  JHEP {\bf 1105}, 085 (2011)
  [arXiv:1103.4279 [hep-ex]].
  %%CITATION = JHEPA,1105,085;%%

%\cite{Alwall:2007st}
\bibitem{Alwall:2007st}
  J.~Alwall, P.~Demin, S.~de Visscher, R.~Frederix, M.~Herquet, F.~Maltoni, T.~Plehn, D.~L.~Rainwater {\it et al.},
  %``MadGraph/MadEvent v4: The New Web Generation,''
  JHEP {\bf 0709}, 028 (2007).
  [arXiv:0706.2334 [hep-ph]].

%\cite{D0}
\bibitem{D0}
  D0 Collaboration, 
  %``Study of the dijet invariant mass distribution in \boldmath{$p\bar{p}\to
  %W(\to\ell\nu)+{jj}$} final states at $\sqrt{s} =1.96$ TeV},''
  arXiv:1106.1921 [hep-ex].
  %%CITATION = ARXIV:1106.1921;%%

%\cite{jointtaskforce}
\bibitem{jointtaskforce}
  CDF, D0, E.~Eichten, and K.~Ellis, 
  \url{http://www.fnal.gov/pub/today/archive_2011/today11-06-10_readmore.html}.

%\cite{Lavoura:1993nq}
\bibitem{Lavoura:1993nq}
  L.~Lavoura and L.~F.~Li,
  %``Making the small oblique parameters large,''
  Phys.\ Rev.\  D {\bf 49}, 1409 (1994)
  [arXiv:hep-ph/9309262].
  %%CITATION = PHRVA,D49,1409;%%

%\cite{Kribs:2007nz}
\bibitem{Kribs:2007nz}
  G.~D.~Kribs, T.~Plehn, M.~Spannowsky and T.~M.~P.~Tait,
  %``Four generations and Higgs physics,''
  Phys.\ Rev.\  D {\bf 76}, 075016 (2007)
  [arXiv:0706.3718 [hep-ph]].
  %%CITATION = PHRVA,D76,075016;%%

\end{thebibliography}
\end{document}